\documentclass[letterpaper]{JHEP3}
\usepackage{epsfig}

\newcommand\fverb{\setbox\pippobox=\hbox\bgroup\verb}
\newcommand\fverbdo{\egroup\medskip\noindent%
			\fbox{\unhbox\pippobox}\ }
\newcommand\fverbit{\egroup\item[\fbox{\unhbox\pippobox}]}
\newbox\pippobox

\title{The Radionactive Universe}

\author{Edward W.\ Kolb \\
         Theoretical Astrophysics Group, Fermi National Accelerator Laboratory,
         Box 500, Batavia, Illinois 60510, and Department of Astronomy and
         Astrophysics, Enrico Fermi Institute, The University of Chicago, 
         Chicago, Illinois 60637 \\
         E-mail: \email{rocky@fnal.gov}}
\author{G\'eraldine Servant \\
         High Energy Physics Division, Argonne National Laboratory, Argonne,
         Illinois 60439, and Enrico Fermi Institute, The University of Chicago,
         Chicago, Illinois 60637\\ E-mail: \email{servant@theory.uchicago.edu}}
\author{Tim M. P. Tait\\
        Fermi National Accelerator Laboratory, Box 500, Batavia, Illinois 
        60510 \\
        E-mail: \email{tait@fnal.gov} }

\preprint{ANL-HEP-PR-03-021\\
EFI-03-12\\
FERMILAB-Pub-03/164-T\&A\\
hep-ph/0306159}

\abstract {In a large class of extra-dimensional models, a scalar
degree of freedom known as the {\em radion} is long-lived, or even
stable, on cosmological scales.  In this paper we investigate the
impact of radionactivity on the evolution of the universe.  We
demonstrate that whether the radion overcloses the universe,
constitutes the dark matter, is the inflaton, the curvaton, or does
not play any role in cosmology, depends crucially on the ratio between
the energy densities stored in the radion and in the inflaton at the
time of inflation.  We discuss the general difficulties reconciling
models with low compactification scale ({\em i.e.,} TeV scale) with
the simple picture of inflation.}


\begin{document}

\section{Introduction}
\label{sec:intro}

Scalar fields are believed to play important roles in both particle
physics and cosmology.  The crucial aspect of scalar fields is that
they determine the properties of the vacuum without relinquishing the
key property: invariance under spacetime symmetries.  In spite of
these very desirable properties, scalar fields fit rather uneasily
within quantum field theories.  Scalar field masses are extremely
sensitive to quantum corrections.  They exhibit quadratic divergences,
which suggests that their natural mass scale should be very large, and
implies that their detailed properties should be controlled by
ultraviolet (UV) physics beyond our ability to infer from current
experiments.  The large masses natural for scalar fields tend to spoil
their potential utility in both particle physics and cosmology.
Idealy, we should be led to consider theories with some mechanism to
protect the scalar mass from large corrections.

In this article we explore one generic feature of theories with extra
spatial dimensions.  Such theories contain a scalar field, the radion,
which plays the role of the geometrical modulus that determines the
size of the compactified dimensions.  From the higher dimensional
perspective, the radion is part of the enlarged graviton tensor, and
at distances smaller than the size of the extra dimension it is
protected from receiving a large mass by the same general covariance
which protects the graviton mass.  Thus, unlike generic scalar fields,
it can be naturally light.  Light scalar degrees of freedom are often
invoked in cosmology: inflatons, curvatons, modulons (fields
controlling coupling strengths), dark matter, {\em etc.} It is
interesting to explore the possibility that the radion may be one of the
much-needed light scalar degrees of freedom, and to see if it can play
an interesting role in cosmology.

Extra-dimensional models with a TeV compactification scale
\cite{Antoniadis:1990ew} have received revived interest lately.  There are 
many ways in which extra dimensions can attempt to explain the outstanding
mysteries in particle physics 
\cite{Arkani-Hamed:1999dc,Arkani-Hamed:2000hv,Antoniadis:1998sd}.  
For example, there is the prospect of dark matter consisting of the
first Kaluza--Klein (KK) mode of the hypercharge gauge boson ($B^1$)
\cite{Servant:2002aq} in a sub-class of these models referred to as 
{\it universal extra dimensions} (UED) \cite{Appelquist:2000nn}.
However, the cosmology of models with TeV size extra dimensions
remains largely unexplored.  Our concern in this paper is to
investigate, for all possible compactification scales, constraints
from radion cosmology on these models, and to a larger extent, on any
models with flat extra dimensions. Different recent attempts
discussing the role of the radion during inflation can be found in
\cite{Fairbairn:2003yx}.

The destiny of a radionactive universe depends on two parameters, the
compactification scale $M_c$ (naturally $L^{-1}$, the physical size of
the extra dimensions) and the ratio $r=M_c/M_I$ where $M_I=V_I^{1/4}$
is the scale of the inflaton potential. One question to ask concerns
the epoch at which the extra dimensions were stabilized compared to
the epoch of inflation. When did compactification take place relative
to inflation?  A first possibility is that inflation took place before
compactification.  This would require a post-inflationary exponential
contraction of extra dimensions, and seems counterintuitive.
Further, it would require that the inflation scale $M_I$ is much
larger than the compactification scale $M_c$.  As is well known, the
non-renormalizability of extra-dimensional theories implies that they
have a low cut-off scale $\Lambda$, beyond which the theory is
strongly coupled and loses predictivity.\footnote{For theories in
which gauge fields propagate in the bulk, na\"{\i}ve analysis suggests
$\Lambda \sim 100 M_c$ (see however
\cite{Chivukula:2003kq} which suggests $\Lambda \sim M_c$) if there are 5
dimensions, and $\Lambda$ is closer to $M_c$ for theories with more dimensions.  The
scale at which gravitational interactions become strong is typically about
$M_5^3\sim M_cm_{Pl}^2 \gg \Lambda^3$ in 5d.}

The second possibility stabilizes the extra dimensions, and then inflates the
four large ones much later.  In this case, geometrical moduli such as the
radion are active during inflation.  These are scalar fields with
Planck-suppressed couplings, and may contribute to the inflaton potential. In
fact, they may be responsible for triggering inflation, for all possible
compactification scales \cite{Arkani-Hamed:1999gq}. Even if they do not have
the properties required to trigger inflation, they may still play a crucial
role in the generation of cosmological perturbations and/or make up the dark
matter.  As we shall see, these scalar fields are typically light, with masses
protected by higher dimensional general covariance.  It is well known that when
the mass is smaller than the Hubble rate during inflation,
the field is frozen during inflation. After inflation, the
energy density stored in these fields can easily overclose the universe. This
is the essence of the moduli problem. Another well known fact is that during
inflation, scalar fields receive corrections to the their mass of order $H$
from non renormalizable operators of the type $ V_I \phi^2 /m^2_{Pl}$ (where
$\phi$ is the modulus in question).  Scalars with masses larger than
approximately $H$ are damped out to the minimum of their potential during
inflation. However, their energy is not necessarily inflated away 
because the position of the minimum is shifted by the same type of operators
which are responsible for shifting the masses. 
As a result, 
energy stored in the radion fields typically survives
inflation and can have dramatic effects on post-inflationary cosmology.  We can
actually predict $m^2/H^2$ during inflation as a function of $r$. We
show that if $r\lesssim 1$, models with a compactification scale $M_c\lesssim
10^9$ GeV can easily be ruled out (at least within a given inflationary
paradigm) because of the overclosure of the universe by radion
oscillations. This means that inflation must take place at a scale as low as
or lower than the compactification scale.

Inflation taking place after compactification appears problematic in models
with a low compactification scale, {\em i.e.,} $M_c \sim$ TeV. This implies
that the highest energy scale of the universe when inflation starts is $M_I
\lesssim M_c$.  From the COBE constraint $\delta_H\sim 10^{-5}$
\cite{Bennett:1996ce}
and by identifying the energy density fluctuations with primordial 
quantum fluctuations of the inflaton, one obtains:
\begin{equation}
\label{COBE}
\left(\frac{V_I}{\epsilon}\right)^{1/4}\sim 10^{16}\ \mbox{GeV} ,
\end{equation}
where $\epsilon=m_{Pl}^2({V_I}^{\prime}/V_I)^2/2$ is the slow-roll parameter.
This data disfavors models with low-scale inflation. Equation~(\ref{COBE})
tells us that if the primordial density perturbations originate from the
inflaton, inflation at the TeV scale requires a slow-roll parameter of the
order of $10^{-48}$, a tiny value which is difficult to motivate naturally.
However, it has recently been hypothesized that structures in the universe may
not necessarily follow from primordial quantum fluctuations in the inflaton,
but could instead be due to another scalar field, referred to as the {\it
curvaton} \cite{Lyth:2001nq}. The curvaton scenario has the advantage that the
constraints on the inflaton potential are relaxed, though at the cost of
introducing another scalar field whose mass must somehow be protected.
Furthermore, the constraints are not relaxed to the point that TeV inflation
becomes natural. Indeed, the COBE constraint in the curvaton scenario
translates into
\begin{equation}
\label{COBE2}
\frac{f_D H_*}{\phi_*}\approx 10^{-4},
\end{equation}
where $H_*$ and $\phi_*$ are the Hubble constant and the amplitude of the
curvaton field respectively at the time of horizon exit. Here $f_D$ is the
fraction of the total energy density contained in the curvaton when it decays
and should be of order unity for this mechanism to work.  To satisfy Eq.\
(\ref{COBE2}), an inflation scale at a TeV would require the small value
$\phi_*\sim 10^{-9}$ GeV $\ll M_I$ for the curvaton amplitude, which is hard to
reconcile with the possibility that the curvaton dominates the total energy
density at the time of its decay. An even more recent proposal
\cite{Dvali:2003em} suggests that cosmological perturbations could be due to
primordial fluctuations in the field (which we will refer to as a {\it
modulon}) whose VEV gives the coupling between the inflaton and the Standard
Model fields and plays a crucial role in the reheating process. In this work we
will investigate whether the radion could ever play the role of a curvaton or a
modulon.

A third alternative has compactification and inflation taking place at
comparable scales.  We assume that the relevant history of our universe starts
at a time when all dimensions have comparable sizes.  Inflation starts in all
$4+n$ dimensions (with a typical scale around or a few orders of magnitude
below $m_{Pl}$).  Later on, the extra dimensions stop inflating, for example
when $M_c$ is at the TeV scale, and the rest of the 4 dimensions continue to
inflate. One assumes there are some extra fields living in the extra dimensions
responsible for the stabilization dynamics and allowing the extra dimensions to
undergo only $15-20$ e-folds of inflation while allowing the other dimensions
to continue to inflate.  In this case, the scale in the inflaton potential is
still much higher than the compactification scale, $M_I
\gg M_c$, and there would be no need for a huge fine-tuning in the epsilon
parameter or in $\phi_*$ to get the right COBE normalization.  As we will see
in section~\ref{sec:radionDM}, there is a serious moduli problem associated
with this scenario. Indeed, when the ratio between the compactification scale
and the inflaton scale is much smaller than one, we are in the regime where
there is no damping at all of the radion field during inflation, and it will
overclose the universe.  Avoiding the overclosure problem typically requires
$M_I \ll M_c$, which on the other hand disfavours low compactification scales
from the COBE constraint, unless there is a new mechanism to generate density
perturbations. Obviously, there are interesting cosmological challenges
associated with TeV compactification scales. In this work, in addition to
examining the difficulties associated with low compactification scales, we will
be able to constrain even larger compactification scales, up to $10^{12}$ GeV.
This leads us to study of more traditional compactifications models, {\em
i.e.,} associated with smaller extra dimensions.

In summary, as soon as compactification took place before inflation, we are
presented with additional scalar fields during inflation, whose masses are
protected from large corrections, and thus they typically survive inflation.
Note that as the number of extra dimensions increases, the number of geometric
moduli also increases. We concentrate on the volume-modulus, {\em i.e.,} we
focus on the 5d case for illustrative purposes. In principle, one may also
consider more extra dimensions, and those moduli which describe the shape 
of the compact manifold.

Section~\ref{sec:radion} reviews the definition and properties of the radion
field and contains a few remarks about the possibility that the radion is
itself the inflaton.  Section 3 stresses the consequences of the generic form
of the radion coupling to any form of energy density and how the radion can be
naturally induced to be an inflaton.  In section~\ref{sec:D} we evaluate the
energy density stored in the radion at the end of inflation and discuss its
evolution when it begins to oscillate.  For small enough $M_c$, the radion is
stable on the time-scale of the age of the universe, and could either account
for the dark matter, or overclose the universe.  This is explored in
section~\ref{sec:radionDM}.  In section~\ref{sec:curvaton}, after studying
 the radion isocurvature
perturbations, we discuss
whether the radion can be a viable curvaton or modulon candidate, and can
account for the observed density perturbations. Section 7 studies the
constraints associated to the late decay (after BBN) of radions and KK
gravitons. In section 8 we summarize the challenges associated with TeV
compactification scales and conclude.

\section{Radion Properties}
\label{sec:radion}

In this section, we briefly review the crucial properties of the radion: its
mass, and its coupling to bulk fields.  When reducing the curvature term of the
($\hat{d}=d+n$)-dimensional action, we obtain terms of the form
\begin{eqnarray}
\label{action}
\nonumber
-\frac{1}{\hat{\kappa}^2}\int 
d^{\hat{d}}x\sqrt{g^{(\hat{d})}}{\cal{R}}^{(\hat{d})}&=&
\int d^dx\sqrt{g^{(d)}} \sqrt{g^{(n)}} e^{\frac{d-2}{2}\Sigma} 
\left[-\frac{1}{\kappa^2}
{\cal{R}}^{(d)} +\frac{1}{4}e^{-\Sigma}g_{ij}F^i_{\mu\nu}F^{j\mu\nu}\right. \\
&-&\frac{(d-1)(d-2)}{4\kappa^2}(\partial_{\mu}\Sigma)(\partial^{\mu}\Sigma) 
-\frac{d-1}{2\kappa^2} g^{ij}(\partial_{\mu}g_{ij})(\partial^{\mu}\Sigma)\\
&+&\left. \frac{1}{4\kappa^2}g^{ij}
(\partial_{\mu}{g_{jk}})g^{kl}(\partial^{\mu}{g_{li}})
-\frac{1}{4\kappa^2}g^{ij}
(\partial_{\mu}{g_{ij}})g^{kl}(\partial^{\mu}{g_{kl}}) + ...\right]
\nonumber
\end{eqnarray}
where the general metric is given by
\begin{equation}
ds^2=(e^{\Sigma} g_{\mu\nu}+ 4g_{ij}\kappa^2A^i_{\mu} A^j_{\nu}) 
dx^{\mu}dx^{\nu} 
+g_{ij} (2\kappa A^i_{\mu}dx^{\mu}dy^j+
2\kappa A^j_{\nu}dx_{\nu}dy^i)+g_{ij}dy^idy^j ~~.
\end{equation}
All fields are functions of both extra-dimensional coordinates $y_i$
and four-dimensional coordinates $x^\mu$.  However, the zero-mode
fields are independent of the $y_i$, and thus we may drop this
dependence when describing the low-energy (zero-mode) physics.  In
five dimensions, the KK tower of the radion is eaten by the massive KK
gravitons leaving only the radion zero mode; in more extra dimensions,
some of the radion KK modes remain as well.  $\hat{\kappa}$ is
Newton's constant in $\hat{d}$ dimensions and is related to Newton's
constant in $d$ (flat) dimensions ($\kappa=m_{Pl}^{-2}$ where $m_{Pl}$ is the
reduced 4-d Planck mass) by $\kappa=\hat{\kappa}/\sqrt{V_n}$ where
$V_n$ is the volume of the compact space. Here $g^{(d)}$ and $g^{(n)}$
are the determinants of the metrics $g_{\mu\nu}$ and $g_{ij}$. In the
right-hand side, $\mu$ and $\nu$ indices go up and down with the
$g_{\mu\nu}$ metric and its inverse.  To obtain a curvature term in
$d$ dimensions which is canonically normalized, we make the following
rescaling:
\begin{equation}
\sqrt{g^{(n)}}e^{\frac{d-2}{2}\Sigma}=1, \ \ \ \  i.e., \ \ \ 
e^{\Sigma}=1/\sqrt{g^{(n)}} ~~ .
\end{equation} 
In this case, Eq.\ (\ref{action}) simplifies to
\begin{eqnarray}
\label{action2}
-\frac{1}{\hat{\kappa}^2}\int d^{\hat{d}}x\sqrt{g^{(\hat{d})}}
{\cal{R}}^{(\hat{d})}&&=
\int d^dx\sqrt{g^{(d)}}  \left[ -\frac{1}{\kappa^2}
{\cal{R}}^{(d)}+\frac{1}{4}{g^{(n)}}^{\frac{1}{d-2}}g_{ij}
F^i_{\mu\nu}F^{j\mu\nu} \right.\\
+&&\left. \frac{1}{4\kappa^2}g^{ij}(\partial_{\mu}{g_{jk}})g^{kl}
(\partial^{\mu}{g_{li}})
+\frac{1}{4(d-2)\kappa^2}g^{ij}(\partial_{\mu}{g_{ij}})g^{kl}
(\partial^{\mu}{g_{kl}}) + ... \right] .
\nonumber
\end{eqnarray}

We now apply these results to the 5-dimensional case, $S^1 / Z_2$ with
bulk fields. The 5d action can be expressed as
\begin{eqnarray}
{\cal L}_5 =  
\sqrt{g} \left\{
M_5^3 R_5 - \frac{1}{4 g_5^2} {\rm Tr}[ F^{M N} F_{M N} ]
- M_c^5 \widetilde{V}_\phi (\tilde{\phi}) 
- \frac{1}{2}M_\psi^3 
\partial_M \tilde{\psi} \partial^M \widetilde{\psi} - 
M_\psi^5 \widetilde{V}_\psi (\tilde{\psi})
\right\},
\end{eqnarray}
where $g$, $M_5$, and $R_5$ refer to the 5d gravitational quantities, $F^{MN}$
is the field strength of a bulk gauge field with 5d coupling $g_5$,
$\tilde{\psi}$ represents a minimally coupled real bulk scalar with potential
$\widetilde{V}_\psi$, and $\widetilde{V}_\phi$ is the radion potential.
For simplicity, we have neglected the possibility
that $\tilde{\psi}$ mixes with the Ricci scalar; we will return to this later.

A convenient parametrization for the 5d metric can be written in the 
form \cite{Appelquist:1983vs}: 
\begin{equation}
\label{metric}
ds^2=\left(e^{-1/3 \tilde{\phi}}g_{\mu\nu} + e^{2/3 \tilde{\phi}}
\widetilde{A}_{\mu} \widetilde{A}_{\nu} \right) 
dx^{\mu}dx^{\nu} 
+2 e^{2/3 \tilde{\phi}}\widetilde{A}_{\mu}dx^{\mu}dy+e^{2/3 \tilde{\phi}}dy^2,
\end{equation}
where $\widetilde{A}_{\mu}=2 \kappa A_{\mu}$ is the gravi-photon field, which
is odd under the orbifold and thus has no zero mode, $g_{\mu \nu}$ is the
tensor graviton field, and $\tilde{\phi}$ is the (dimensionless) gravi-scalar
(radion).  This form is a convenient starting point because it decouples the
kinetic terms of the radion and the transverse tensor modes and produces a
radion kinetic term of the canonical form.

To determine the effective action at scales less than of order $1/L$, we
replace the 5d fields with their zero-mode components, and integrate out the
extra dimension.  In doing so, it is convenient to introduce a reference length
$L$.  Note that the physical size of extra dimensions is
$L_{phys}=e^{\tilde{\phi}/3} L$.  The effective action for the zero modes is,
\begin{eqnarray}
{\cal L}_4 & = &
\sqrt{g} \left[ M_5^3 L R_4 
- \frac{M_5^3 L}{6} \partial_\mu \tilde{\phi} \partial^\mu \tilde{\phi}
- \frac{L}{4 g_5^2} e^{\tilde{\phi}/3} F_{\mu \nu} F^{\mu \nu}
- \frac{M_\psi^3 L}{2} \partial_\mu \tilde{\psi} \partial^\mu \tilde{\psi} 
\right. \nonumber \\ & & \left. \phantom{\frac{M_^^3}{3}}
- M_\psi^5 L e^{- \tilde{\phi}/ 3} \widetilde{V}_\psi (\tilde{\psi})
- M_c^5 L \widetilde{V}_\phi(\tilde{\phi}) \right] ,
\label{effectiveaction}
\end{eqnarray}
where now all quantities refer to the 4d (zero-mode) fields.  From here it is
simple to rescale fields to canonical normalization and define four-dimensional
scalar potentials,
\begin{eqnarray}
\label{eq:normalize}
\phi & = & \sqrt{M_5^3 L/3}\ \tilde{\phi}, \qquad 
\psi = \sqrt{M_\psi^3L}\ \tilde{\psi}, \cr
V_\phi(\phi) & = & LM_c^5\widetilde{V}_\phi(\tilde{\phi}), \qquad
V_\psi(\psi)   =   LM_\psi^5\widetilde{V}_\psi(\tilde{\psi}).
\end{eqnarray}

In order for the low energy physics to appear four dimensional, the radion
potential must provide a stable VEV for $\phi$.  We can adjust the reference
quantity $L$ such that this happens for $\langle \phi \rangle = 0$.  Clearly,
unless there is fine-tuning in the radion potential, this will occur for $L
\sim M_c^{-1}$.  For this choice, $L$ is in fact the physical size of the extra
dimension, and we can identify $M_5^3 L = m_{Pl}^2$ and $g_5^2 / L = g_4^2$.
The radion couplings are thus,
\begin{eqnarray}
\label{radioncouplings}
- \frac{1}{4} e^{\phi/\sqrt{3} m_{Pl}} F_{\mu \nu} F^{\mu \nu}
- e^{-\phi/\sqrt{3}m_{Pl}} V_\psi (\psi),
\end{eqnarray}
where we have rescaled the kinetic terms for the bulk fields
to canonical normalization as in Eq.\ (\ref{eq:normalize}).

\subsection{Radion Mass}

In the absence of contributions from $V_\psi$, which we consider in the
next section, the radion mass may be determined by expanding the potential 
$\widetilde{V}_\phi$ appearing in Eq.~(\ref{effectiveaction}) 
to order $\tilde{\phi}^2$
about its minimum,
\begin{eqnarray}
- M_c^5L \frac{1}{2} \widetilde{V}_\phi^{\prime \prime}(\tilde{\phi})
\tilde{\phi}^2 + \ldots 
= - M_c^5L \frac{1}{2} \widetilde{V}_\phi^{\prime \prime}(\tilde{\phi}) 
\frac{3}{m_{Pl}^2}\phi^2 + \ldots
\end{eqnarray}
where primes refer to functional derivatives with respect to $\tilde{\phi}$.
Without fine tuning, one expects $L \sim M_c^{-1}$, and all derivatives of
$\widetilde{V}_\phi$ to be order unity.  Thus, the radion mass is naturally
\begin{equation}
\label{radionmass}
m_r \sim \sqrt{3}\frac{M_c^2}{m_{Pl}} .
\end{equation}
Note that loop corrections will not destabilize the radion mass, as they are of
order ${M_c^2}/{m_{Pl}}$. Indeed they are proportional to the radion coupling
and they must involve $M_c$ since in the limit of decompactification the
radion is part of the higher dimensional graviton, and thus is massless.  The
fact that the radion is typically light and its mass protected was stressed in
\cite{Chacko:2002sb} where it was also argued that for $M_c \sim$ TeV, the
radion would be responsible for detectable deviations to Newtonian gravity in
present or future experiments.  Thus, there is a lower bound on $M_c$ from
short-range gravitational 
experiments \cite{Chacko:2002sb,Antoniadis:2002gw,Perivolaropoulos:2002pn}.
A radion with a mass of $10^{-3}$ eV would induce modification of Newton's law
at distances of order $100\,\mu$m (at the edge of distances probed by
gravitational experiments). No deviation has been measured, therefore
\begin{equation}
m_r\gtrsim 10^{-3} \ \ \mbox{eV} \ \rightarrow M_c\gtrsim 0.8 \ \mbox{TeV} ,
\end{equation}
assuming the gluons are bulk fields.  The limits for brane gluons are weaker
as a result of a weaker coupling between brane nucleons and the radion 
(see below).  In these cases, direct collider limits will provide
a lower bound on $M_c$ of order 1 TeV, the precise value depending on
the extra-dimensional framework \cite{Appelquist:2002wb}.

\subsection{Radion Couplings}

Radion couplings to bulk fields may be simply read from
Eq.~(\ref{radioncouplings}), and are characterized by the four dimensional
Planck scale, $m_{Pl}$.  For much of the parameter space, $M_c \lesssim 10^8$
GeV, the mass of the radion is less than 1 MeV, and decays proceed only into
photons (and zero-mode gravitons).  These results formally apply only to flat
extra dimensions and bulk SM fields.  Two other possibilities are 
brane world models \cite{Arkani-Hamed:1998rs} in flat extra
dimensions, or warped (RS)
extra dimensions \cite{Randall:1999ee}. In the RS case, 
the radion mass and couplings
are both characterized by the TeV scale 
\cite{Goldberger:1999uk,Csaki:2000zn}.  This large mass
will ruin most of the cosmological applications of the radion we are
considering, and thus we will not consider the warped case any further (see
however, Ref\ \cite{Csaki:1999mp}).

For flat extra dimensions with SM fields on a brane, the radion couplings
are substantially different because brane fields do not contribute to the fifth
components of the stress-energy tensor, and thus couplings to radions are
induced only through the mixing of $g_{55}$ with longitudinal gravitons
\cite{Giudice:2000av}.  The net result is that the radion couples only to the
trace of the brane stress-energy tensor.  For gauge fields, this occurs at
tree-level proportionally to the gauge-field mass, and at one loop
proportionally to the $\beta$ function.  The brane photon thus couples to the
radion more weakly than a bulk photon would, with a relative coupling that is
roughly $16 \pi^2$ smaller.  We parameterize the radion coupling to SM fields
by $c/m_{Pl}$, with $c=1$ for the bulk SM, and $c \sim 10^{-3}$ for the brane
SM case.  However, zero-mode gravitons, as bulk fields, will always couple with
full strength to the radion, and thus the brane photon case does not
substantially affect the radion lifetime, but instead suppresses the branching
ratio for decay into photons.  We will focus on the bulk photon case below, but
occasionally comment on the brane-world scenario where appropriate.

\subsection{Radion Lifetime}

The decay into two bulk photons implies a radion decay width
\begin{equation}
\Gamma = \tau^{-1} \simeq \frac{m_r^3}{192 \pi m_{Pl}^2} \sim
\frac{\sqrt{3} M_c^6}{64 \pi m_{Pl}^5} , 
\label{lifetime}
\end{equation}
where we have ignored a similar contribution from the decay into zero-mode
gravitons.  For larger $M_c$, decays into other SM fields such as light
fermions and hadrons can increase the width by a factor of order unity.  The
radion is effectively stable provided $\tau$ is larger than $H_{0}^{-1}\sim 4
\times 10^{17}$s, corresponding to
\begin{equation}
\label{stability}
M_c\lesssim 7 \times 10^8 \mbox{GeV} .
\end{equation}

\subsection{Radion as the Inflaton}
\label{subsec:radinflaton}

The radion potential at leading order is $V_\phi(\phi) \sim m_r^2\phi^2/2$,
for small $\phi/m_{Pl}$.  At large $\phi/m_{Pl}$ this form represents an
assumption about the form of the $V_\phi(\phi)$. In
order for the radion to be the inflaton, the potential energy density of the
radion must dominate the energy density of the universe. For the radion to be a
viable inflaton candidate, it must satisfy the slow-roll condition. The
quadratic potential is of the {\it chaotic} type, so 
the slow-roll condition does not
depend on the radion mass, but only on the amplitude of $\phi$.  The slow roll
condition can be stated as a condition on a parameter $\epsilon$:
\begin{equation}
\epsilon=\frac{1}{2}m^2_{Pl}\left(\frac{V^{\prime}}{V}\right)^2 
= 2 \frac{m^2_{Pl}}{\phi^2} < 1,
\end{equation}
which is satisfied for $\phi\gtrsim \sqrt{2}m_{Pl}$.  As discussed in the next
section, this condition turns out to be natural for the radion.  Note that the
requirement $\phi\gtrsim m_{Pl}$ is still consistent with a regime where our
effective theory is under control.  Indeed, the mass of the radion being
typically much smaller than $m_{Pl}$, the energy density associated with the
radion is still much smaller than $m_{Pl}^4$, and therefore we are still away
from a regime where 4d quantum gravity corrections would dominate.

One important consideration on inflation is the requirement that the inflaton
energy density be converted to radiation after inflation and prior to
nucleosynthesis and/or the freeze-out of dark matter relics.  The reheat
temperature in a scenario with perturbative reheating is given by
$T_{RH}\sim\sqrt{\Gamma m_{Pl}}\sim M_c^3/m_{Pl}^2$.  We present in Fig.\
\ref{fig:0} the plot of the decay temperature of the radion as a function of
$M_c$, ({\em i.e.,} the reheat temperature of the universe in the scenario
where the radion is the inflaton). Requiring that the reheat temperature is
larger than the TeV scale (so that the electroweak candidates for cold dark
matter such as LSPs and/or LKPs are preserved) leads to the requirement
$M_c\gtrsim 10^{13}$ GeV.

The fact that the radion couples gravitationally to any field, and in particular
to other scalars of the theory, modifies the na\"{i}ve radion scalar potential
written above. The effect is not negligible if the energy of these scalar
fields is significant and leads to an interesting inflationary cosmology as we
discuss now.

\EPSFIGURE[t]{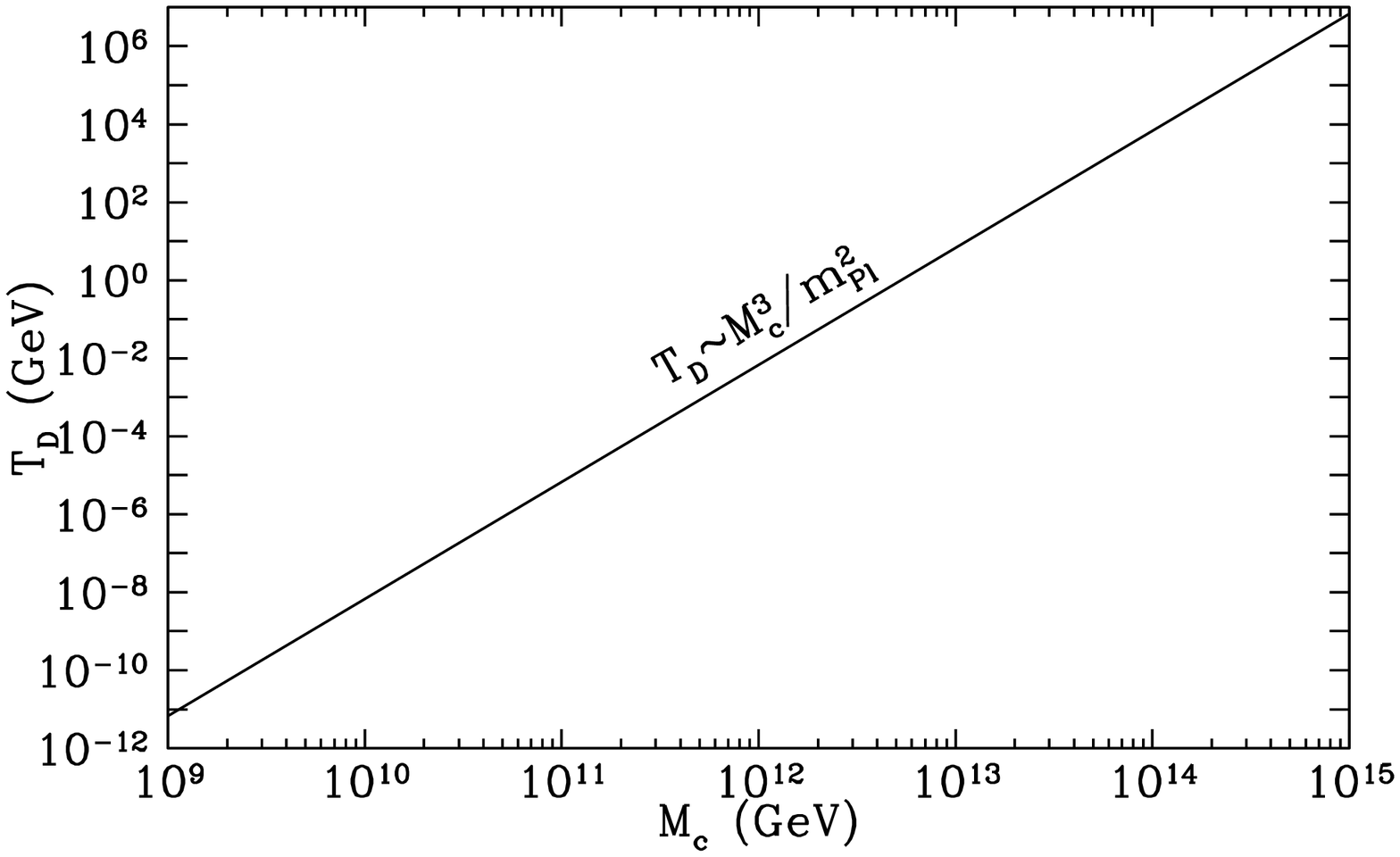,width=15cm}
{Reheat temperature associated with radion decay as a function of the
 compactification scale.\label{fig:0} }

\section{Radion Mixing and Induced Inflationary Potential}
\label{sec:radmixing}

Let us assume the existence of a scalar field $\psi$ other than the radion. If
$\psi$ is a bulk field, according to Eq.\ (\ref{effectiveaction}), the total
scalar potential is
\begin{equation}
V(\phi,\psi)=V_\phi(\phi)+e^{-\phi/\sqrt{3}m_{Pl}} V_\psi(\psi) .
\label{totalpotential}
\end{equation}
We continue to assume that the form
$V_\phi(\phi)=m_r^2\phi^2/2$ remains valid at high energies.  This provides the
simplest phenomenological description of stabilized extra dimensions during
inflation. Even if the position of the minimum is displaced because of the
second term in Eq.\ (\ref{totalpotential}) and thus the size of the extra
dimension is modified, $V_\phi(\phi)\sim m_r^2\phi^2$ guarantees that the VEV
of $\phi$ remains finite and is not destabilized during inflation.
This potential is plotted as a function of $\phi$ for fixed $\psi$ and a few
choices of $r$ in Fig.~\ref{fig:vtotal}.

When $\psi$ is localized on the 4-d brane, the potential is of a slightly
different form:
\begin{equation}
V(\phi,\psi)=V_\phi(\phi)+Ae^{-2\phi/\sqrt{3}m_{Pl}}V_\psi(\psi) .
\label{totalpotential2}
\end{equation}
In this case the coupling between the radion and $\psi$ comes from
$\sqrt{g}_{\mbox{\tiny{induced}}} T^{\mu}_{\ \ \mu}$. Here $A $ is an affine
function of $\xi$, where $\xi$ reflects the presence of terms in the (brane)
action of the type
\begin{equation} 
\sqrt{g}_{\mbox{\tiny{induced}}} \ \xi \ {\cal R} \ (\psi^2 + ...)  
\end{equation}
which mixes the Ricci scalar with $\psi^2$.
One can suppress the $\phi-\psi$ coupling at tree level by suitably choosing
$\xi$, however this coupling will generically reappear at loop
level. Consequently, we expect a potential of the type in Eq.\
(\ref{totalpotential}), and the bulk and brane cases should not be
qualitatively very different.  We will focus on the bulk case below.

We now investigate the virtues of the potential of Eq.\ (\ref{totalpotential})
for inflation.  Because of the $\phi-\psi$ coupling, the VEV of $\phi$ is
shifted from zero and satisfies the relation
\begin{equation}
\langle \phi\rangle= \frac{V_{\psi}}{\sqrt{3}m_r^2}e^{-\langle
\phi\rangle/\sqrt{3}} ,
\end{equation}
where from now on, we refer to $\phi$ (and $\psi$) in Planck units.  We can
easily evaluate $\langle \phi\rangle$ in the case where $V_{\psi}$ is a
constant; for instance, in the regime where $\psi$ is slowly rolling. We
define the scale $M_I$ to be $V_\psi(\psi)\equiv M_I^4$, and
in terms of the parameter $r$ defined as
\begin{equation}
\label{rdefinition}
r \equiv \frac{M_c}{M_I},
\end{equation}
we can express $\langle \phi \rangle$ as
\begin{equation}
\langle \phi\rangle= \frac{e^{-\langle\phi\rangle/\sqrt{3}}}{3\sqrt{3}r^4}
\ \ \mbox{and} \ \ 
V(\langle \phi\rangle)= 3M_I^4r^4\left(\frac{\langle\phi\rangle^2}{2}
+\sqrt{3}\langle\phi\rangle\right) .
\end{equation}
We plot in Fig.\ \ref{fig:phivev} the VEV of $\phi$ as a function of $r$
under these conditions.  

We can see that in the limit $r\gg 1$, the position of the high energy
minimum nearly coincides with the position at low energy ($\phi=0$). When $r
\gg 1$ we have $\langle\phi\rangle \sim 1/(3\sqrt{3}r^4)$ leading to 
$V(\langle \phi\rangle) \approx M_I^4$. Therefore, we recover
$\langle\phi\rangle\approx 0$ and the $\phi-\psi$ coupling is irrelevant. As
can be seen in Fig.\ \ref{fig:potentials}, when $r\gg 1$, $\psi$ dominates the potential and
plays the role of the inflaton.

In the opposite limit, $r\ll 1$, we find $\langle\phi\rangle\gg 1$ (see Fig.\
\ref{fig:phivev}), which leads to a suppression of the contribution of $\psi$
in the potential. When $r \ll 1$, $\phi $ dominates the potential, $V(\langle
\phi\rangle) \approx 3M_I^4r^4\langle\phi\rangle^2/2$, and $\phi$ plays the
role of the inflaton even though $V_{\psi}\gg V_{\phi}$.

\EPSFIGURE[t]{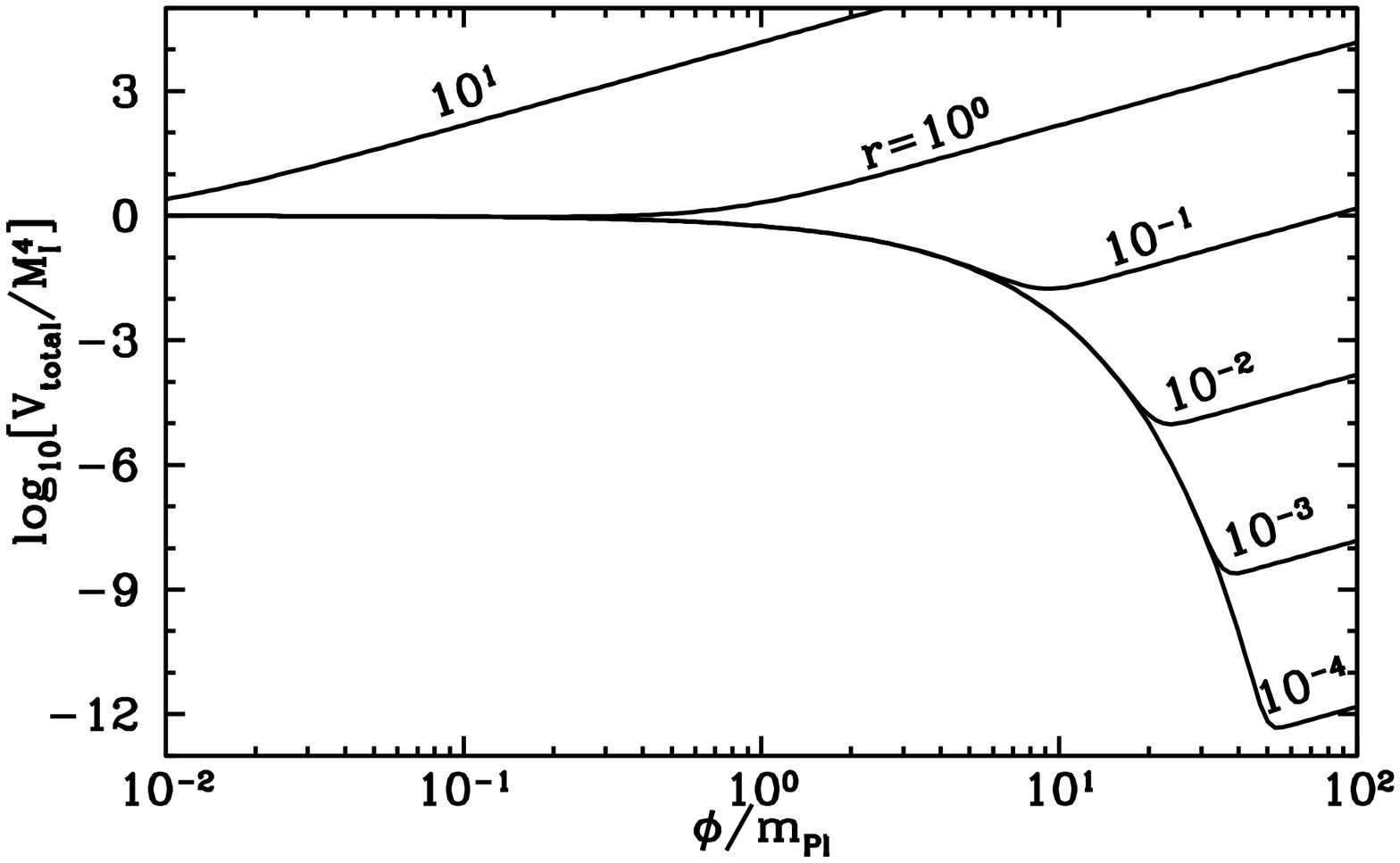,width=15cm}
{The scalar potential as a function of $\phi$ (in Planck units) for
$r=10^1,\ 10^0,\ 10^{-1},\ 10^{-2},\ 10^{-3},\ 10^{-4}$ (from left to right).
 $r$ is defined in Eq.~\ref{rdefinition}.
 \label{fig:vtotal} }

\EPSFIGURE[t]{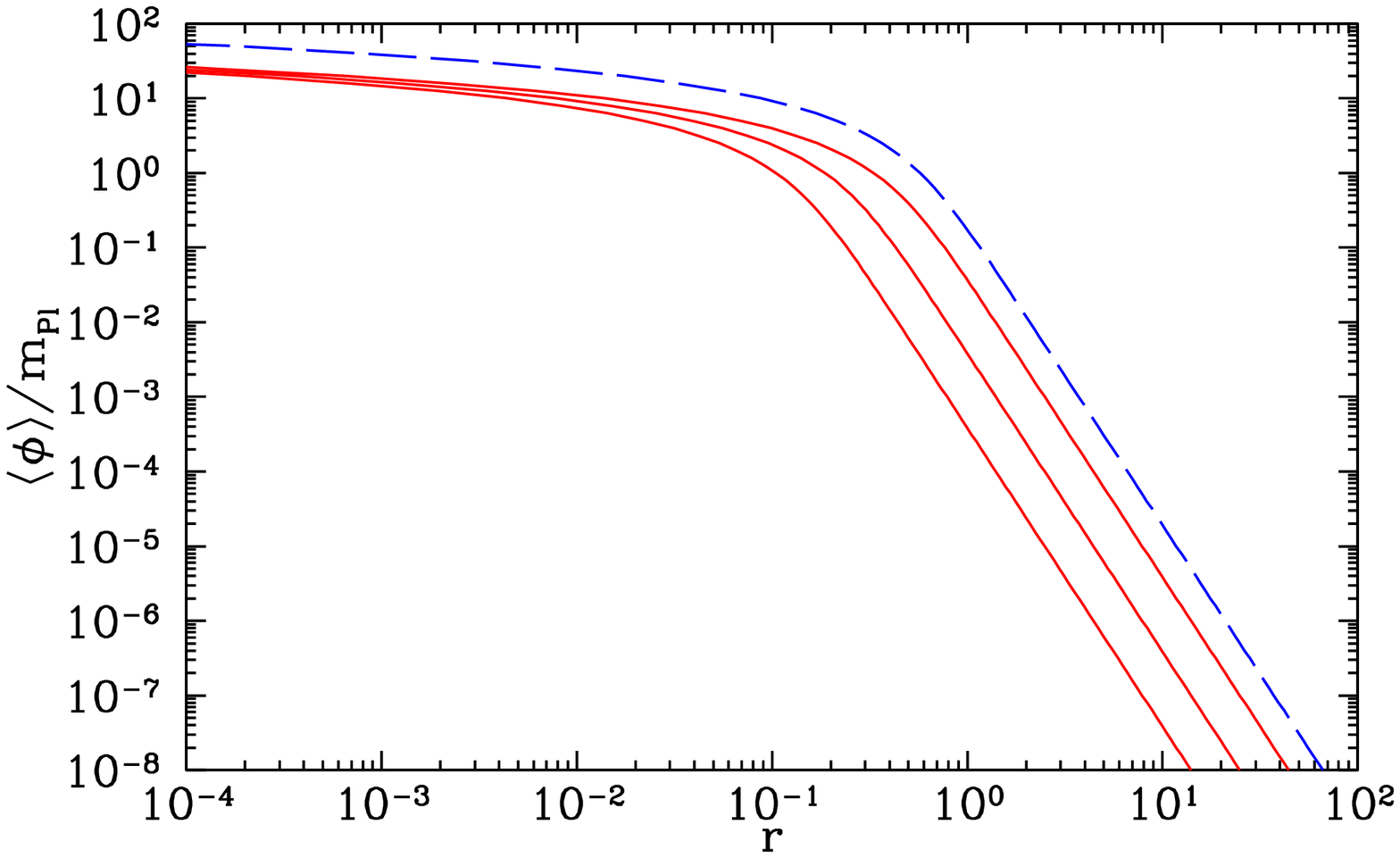,width=15cm}
{The vacuum expectation value of $\phi$ (in Planck units) at the minimum of its
potential as a function of $r$ (as defined in Eq.~\ref{rdefinition}).  The dashed line
 corresponds to the case where
the inflaton lives in the bulk. The plain lines correspond to cases where the
inflaton is localized in four dimensions with suppressed coupling to the radion
parametrized by $A$ as defined in Eq.~(\ref{totalpotential2}). From top to
bottom, the solid curves are for $A=0.1$, 0.01, 0.001.  \label{fig:phivev} }


\EPSFIGURE[t]{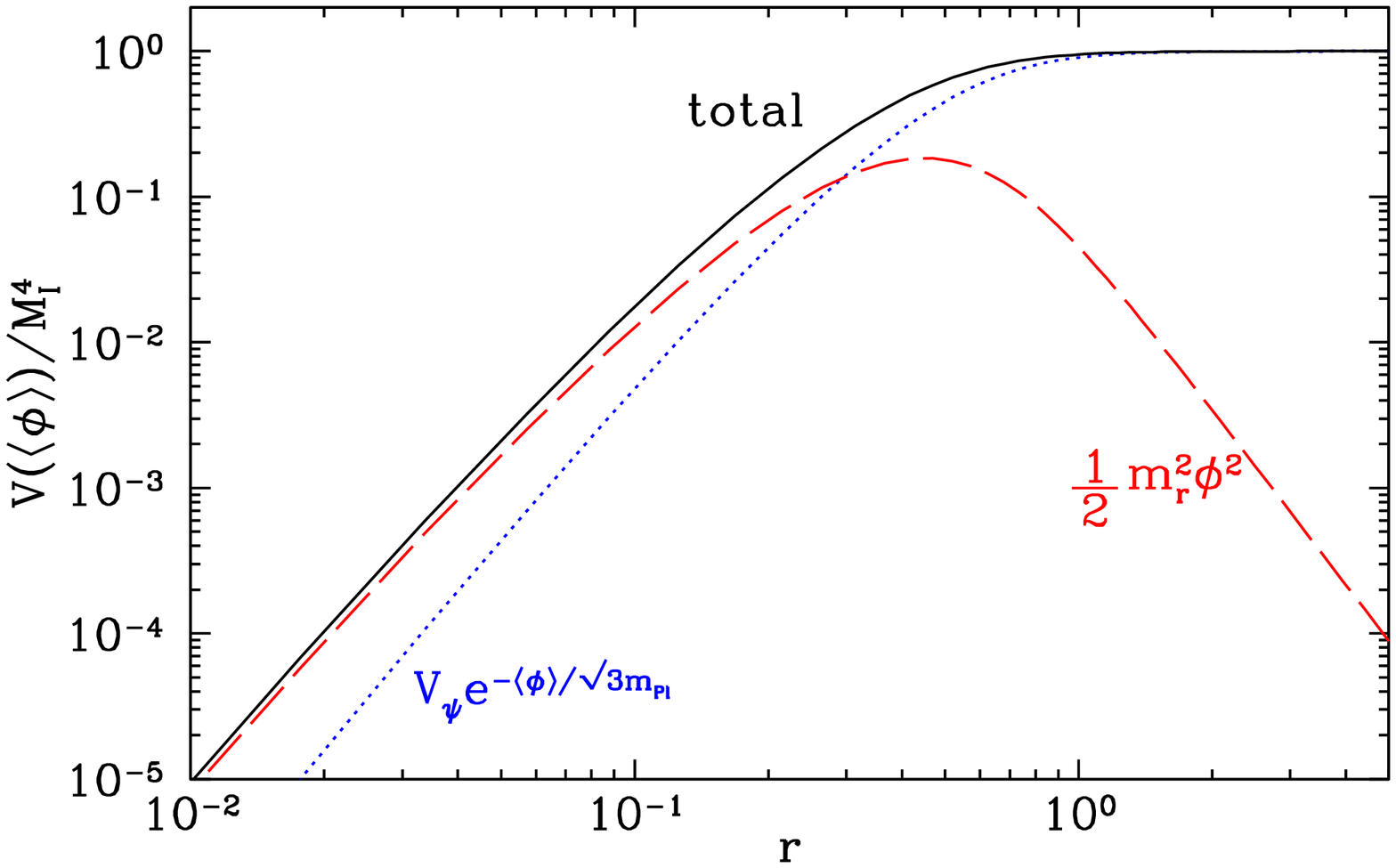,width=15cm}
{Scalar potential (in units of $M_I^4$) evaluated at the minimum of $\phi$
(solid curve). The dashed curve is the contribution from $V_{\phi}$ only. The
dotted curve is the contribution from $\psi$:
$V_{\psi}e^{-\phi/\sqrt{3}}$.  The potential involving the field
$\psi$ dominates the energy density for $r\gg 1$, while for $r\ll 1$, $\phi$
dominates. $r$ is defined in Eq.~\ref{rdefinition}.
\label{fig:potentials} }

A very interesting property of this potential is that even if initially the
conditions were not satisfied for $\phi$ to be an inflaton, $\phi$ will be
driven to be an inflaton through its interactions with $\psi$. If $\psi$
carries enough energy density, it inevitably pushes the radion away from
$\phi=0$ to large values of $\phi$, where the radion can act as an
inflaton. Somewhat counter-intuitively, the higher the energy density of
$\psi$, the more efficiently the radion becomes an inflaton. Because of the
exponential coupling between $\phi$ and $\psi$, the higher $M_I$, the larger
the shift of $\phi$, and the larger the eventual suppression of the energy
density of $\psi$.  Note that during inflation the size of the extra dimension
is modified.  It is larger by a factor $e^{\langle\phi\rangle/{3}}$ compared to
the low energy case. The mass of the radion is also modified to
\begin{equation}
\frac{\partial^2V}{\partial \phi^2}=m_r^2+
\frac{V_{\psi}}{3m^2_{Pl}}e^{-\phi/\sqrt{3}} \ \ \rightarrow \ \ 
m_{\rm{eff}}^2=m_r^2\left(1+\frac{\langle\phi\rangle}{\sqrt{3}}\right) .
\label{effectivemass}
\end{equation}
It is informative to compute the value of the ratio
$m_{\rm{eff}}^2/H^2$ during inflation:
\begin{equation}
\frac{m_{\rm{eff}}^2}{H^2}=\frac{6}{\langle\phi\rangle^2}
\frac{1+\langle\phi\rangle/\sqrt{3}}{1+2\sqrt{3}/\langle\phi\rangle} .
\end{equation}
In the limit $r\ll 1$, ${m_{\rm{eff}}^2}/{H^2}\rightarrow
{2\sqrt{3}}/{\langle\phi\rangle}\ll 1 $. As is well known, in this case the
primordial fluctuations of $\phi$ are frozen during inflation. This confirms
the fact that the radion is a viable inflaton candidate when $r\ll 1$ since
it can be the origin of cosmological perturbations. However, in the opposite
limit $r\gg 1$, the effective radion mass is too large:
${m_{\rm{eff}}^2}/{H^2}\rightarrow {\sqrt{3}}/{\langle\phi\rangle}\gg 1$.

\section{Evolution of the Radion Field}
\label{sec:D}

\subsection{Evolution During Inflation}

We are now interested in determining the evolution of the radion field
in the primordial stages of the universe. From the discussion in the
previous section, one cannot ignore the coupling between $\phi$ and
$\psi$. If we want to follow the cosmological evolution of the radion,
it is necessary to solve the system of coupled differential equations
involving $\phi$ and $\psi$. However, for pedagogical purposes, we
start with the textbook case and first solve the equation of motion for
the VEV of $\phi$ in the absence of such a coupling :
\begin{equation}
\ddot{\phi}+3H\dot{\phi}+V_\phi^{\prime}(\phi)=0 ,
\end{equation}
where $H=M_I^2/\sqrt{3}m_{Pl}$ is a constant during inflation.
In terms of $\tilde{t}=Ht$ and $x=m_r/H=3r^2$, the equation of motion reads 
\begin{equation}
\ddot{\phi}+3\dot{\phi}+x^2\phi=0 ,
\end{equation}
which has solution
\begin{eqnarray}
\phi(\tilde{t})=\frac{e^{-3\tilde{t}/2}}{2\Delta}
\left[(3\phi_0+2\phi_1)\left( e^{\Delta \tilde{t}/2}
                            - e^{-\Delta\tilde{t}/2} \right)
+\Delta\phi_0\left( e^{\Delta\tilde{t}/2}
                  + e^{-\Delta\tilde{t}/2}\right)\right] ,
\end{eqnarray}
where $\Delta=\sqrt{9-4x^2}$, $\phi_0=\phi(\tilde{t}=0)$,
$\phi_1=\dot{\phi}(\tilde{t}=0)$. We typically expect $\phi_1\sim x
\phi_0$ (assuming that the initial energy density of $\phi$ is equally
distributed between kinetic and potential energy). As is well known,
for $x\gg 1$, $\phi$ is exponentially damped during inflation while
for $x\ll 1$, $\phi$ is frozen and remains unaffected by inflation.
However, as we will see now, these conclusions are substantially
altered when including a coupling between the moduli and the inflaton,
even as weak as a gravitational coupling.

If $\psi$ is a bulk field, the coupled equations to be solved are
\begin{eqnarray}
\ddot{\phi}+\sqrt{3}\dot{\phi}\left\{\frac{\dot{\phi}^2}{2}+
\frac{\dot{\psi}^2}{2}+c_r\frac{\phi^2}{2}+\frac{V_{\psi}
e^{-\phi/\sqrt{3}}}{{\cal M}^2m_{Pl}^2}
\right\}^{1/2}+ d_r \dot{\phi} +c_r \phi -
\frac{V_{\psi}{e^{-\phi/\sqrt{3}}}}{\sqrt{3}{\cal M}^2m_{Pl}^2}=0\\
\ddot{\psi}+\sqrt{3}\dot{\psi}\left\{\frac{\dot{\phi}^2}{2}+
\frac{\dot{\psi}^2}{2}+c_r\frac{\phi^2}{2}+\frac{V_{\psi}e^{-\phi/\sqrt{3}}}
{{\cal M}^2m_{Pl}^2}\right\}^{1/2}+ d_i \dot{\psi} +
\frac{V^{\prime}_{\psi}(\psi){e^{-\phi/\sqrt{3}}}}{m_{Pl}{\cal M}^2}=0 ,
\nonumber
\end{eqnarray}
where $\phi$ and $\psi$ are in Planck units, we have defined a dimensionless
time variable $\tilde{t}$ in terms of a convenient mass scale ${\cal M}$ as
$\tilde{t}={\cal M} t$, and we have introduced the constants
\begin{equation}
c_r=\frac{m_r^2}{{\cal M}^2} \ , \ \ \ 
d_r=\frac{\Gamma_{\phi}}{\cal M} \ , \ \
d_i=\frac{\Gamma_{\psi}}{\cal M}.
\end{equation}
Note that the addition of the decay rate terms in the equations of motion is a
an effective description which is valid only when oscillations of the fields
have started \cite{Kofman:1997yn}.  

\begin{figure}
\begin{center}
\epsfig{file=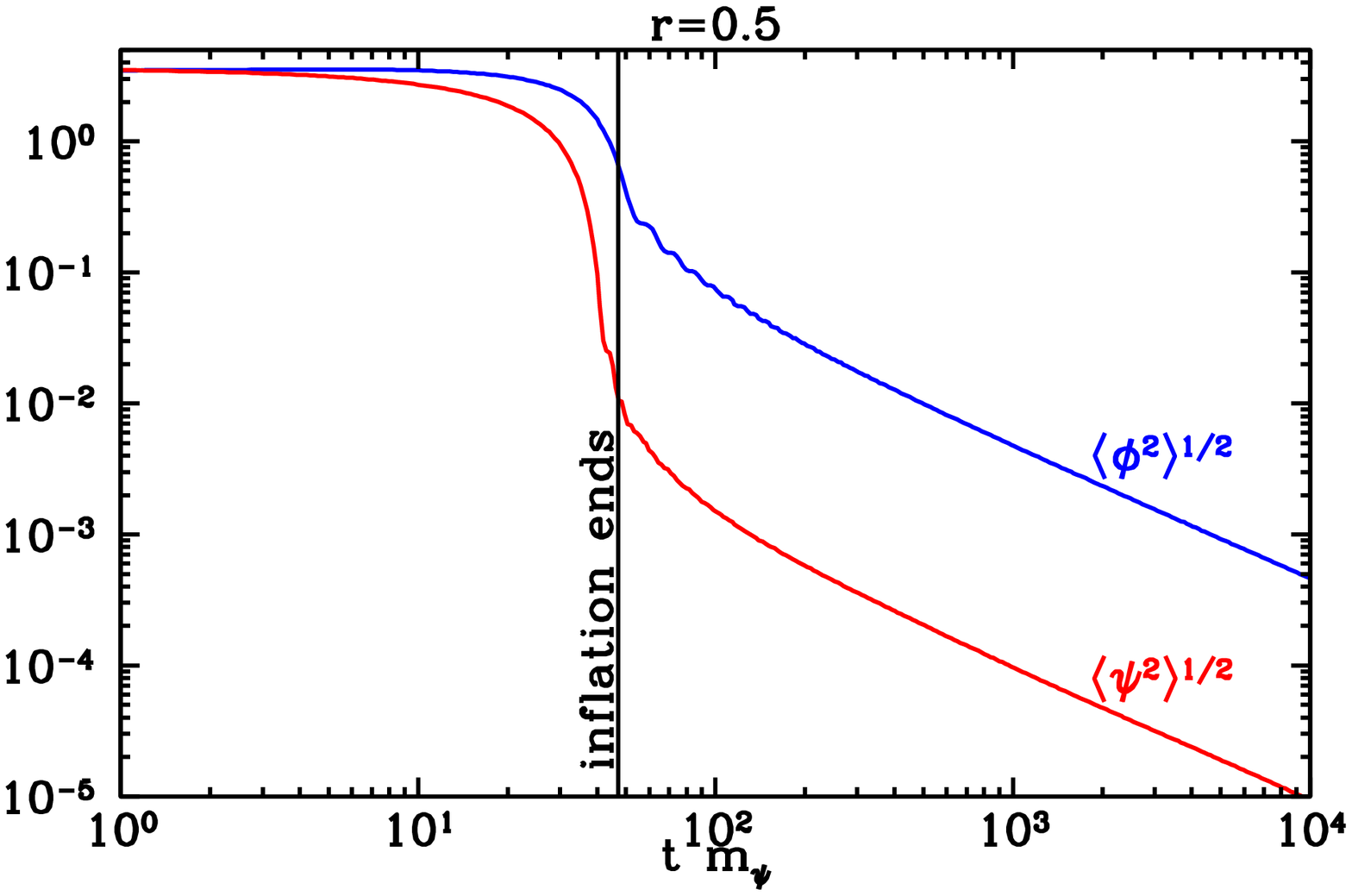,width=12cm} \\
\epsfig{file=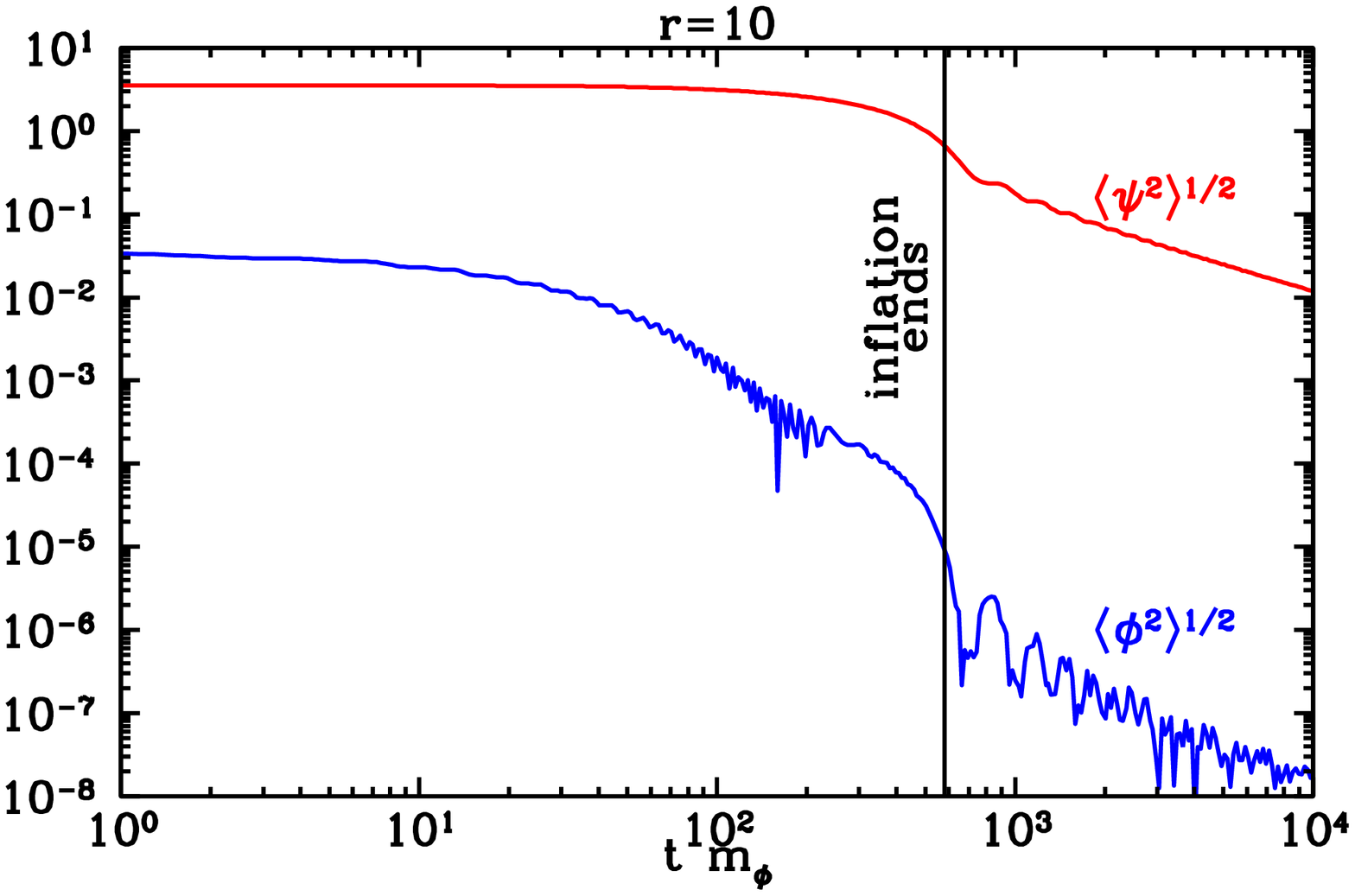,width=12cm} 
\caption{Examples of the time evolution of $\phi$ and $\psi$ using the
potential for $\psi$ of the form $V_{\psi}(\psi)=m_\psi^2\psi^2/2$.  In these
specific examples, $r=0.5$ in the top panel and $r=10$ in the bottom panel
($r=\sqrt{m_r/m_\psi}$). Initial conditions were chosen such that $\psi_0=5$,
and $\phi_0$ to satisfy the equipartition condition. \label{fig:example} }
\end{center}
\end{figure}

We define $D$ to be the value of $\phi$ in Planck units when the field
oscillates and its energy density starts to decrease as a
nonrelativistic component:
\begin{equation}
D \equiv \frac{\phi(t_{osc})}{m_{Pl}}.
\end{equation}
The factor $D$ may be interpreted as the damping factor if the initial value of
the radion is Planckian.  Our goal is to evaluate by which amount the radion
field can be damped during inflation for a given value of $r$, so that we can
investigate the resulting radion energy density today.

\EPSFIGURE[t]{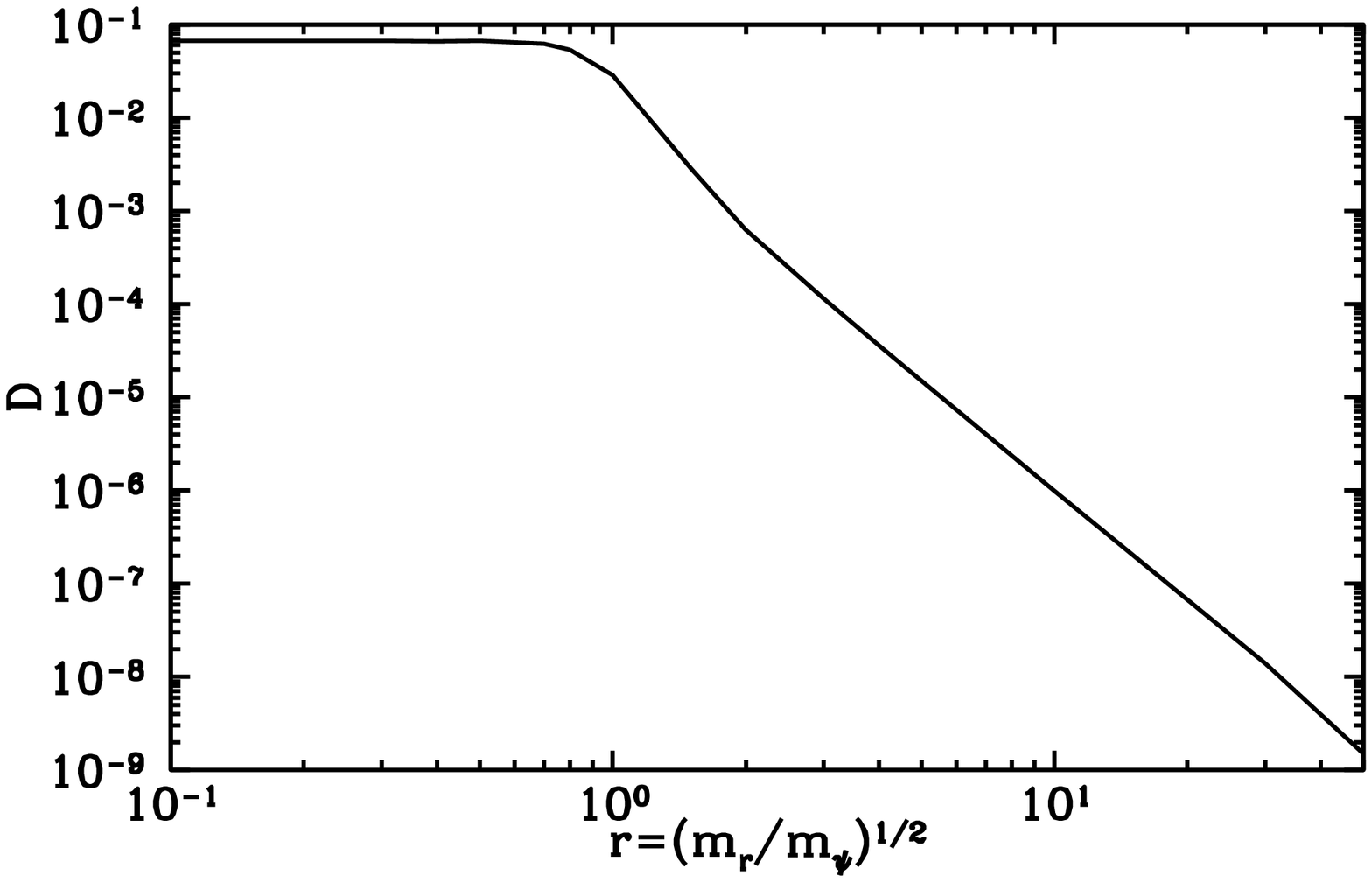,width=15cm}
{$D(r)$ as derived in the chaotic case.  For initial conditions we have assumed
$\psi_0=5$, and have chosen $\phi_0$ to satisfy the equipartition condition.
\label{fig:Dofr}}

The damping of the amplitude of $\phi$ generally depends on the form of the
potential for $\psi$.  However, some general statements can be made. Indeed,
what typically happens can be easily understood by inspection of Fig.\
\ref{fig:phivev}: Because of the coupling between $\phi$ and $\psi$, the
position of the minimum for $\phi$ is displaced from its low energy
position. During inflation, $\phi$ is pushed to its high-energy
minimum and we naturally expect that once inflation ends, oscillations
in $\phi$ start with an amplitude of the order of the VEV of $\phi$ as
given in Fig.\ \ref{fig:phivev}. We checked that this is indeed what
typically happens using different types of potentials for $\psi$.

As an illustration, we show in Fig.\ \ref{fig:example} the time evolution for
the $\phi$ and $\psi$ fields in the case where the potential for $\psi$ is of
the large-field (chaotic inflation) type.\footnote{An interesting possibility
for the origin of the ``precursor'' inflaton $\psi$ is $A_5$, the higher
component of a gauge field
\cite{Arkani-Hamed:2003wu}. }  
What is shown in the figures are the values of
$\langle\phi^2\rangle^{1/2}$ and $\langle\psi^2\rangle^{1/2}$:
\begin{eqnarray}
\langle \phi^2 \rangle^{1/2} & \equiv & \sqrt{\rho_{\phi} / m_r^2} \\
\langle \psi^2 \rangle^{1/2} & \equiv & 
\sqrt{\rho_\psi/m_\psi^2\exp(-\phi/\sqrt{3})},
\end{eqnarray}
where the energy densities capture both the kinetic and potential
energies.  During inflation, $\langle \phi ^2
\rangle^{1/2}=\phi/\sqrt{2}$ and $\langle \psi ^2 \rangle^{1/2}=
\psi e^{-\phi/2\sqrt{3}}/2$. After inflation, when the fields oscillate, the 
averages defined above correspond to cycle averages of the field.

Note that in the case of the chaotic potential
$m_{\psi}^2\psi^2/2$, $r\sim \sqrt{m_r/m_{\psi}}$. For $r$ to be larger than
unity would require that the mass of $\psi$ be somehow less than the radion
itself.  This may be difficult to achieve naturally, however, we consider 
this possibility since we find that the chaotic potential for $\psi$ 
provides the largest damping during inflation. 
(See the $D(r)$ plot derived in this case in Fig.\ \ref{fig:Dofr}). 
As initial conditions, one may choose equipartition of the
potential energy densities between $\phi$ and $\psi$, leading to
\begin{equation}
e^{-\phi_0/\sqrt{3}} = r^4\phi_0^2 /\psi_0^2,
\label{equip} 
\end{equation}
and the initial condition for $\psi_0$ has to be chosen such that 
$\psi_0$ satisfies a slow-roll condition.

In the following sections we will express cosmological constraints as
general bounds on $D$, and will illustrate how these constraints typically
translate into $r$, remembering that for $r<1$, we expect $D$ will  
naturally be of order unity.

\subsection{Validity of 4D effective field theory}

Throughout the discussion we have assumed an effective 4-dimensional
description. The 4d effective field theory is an approximation which neglects the effects
of higher KK modes. Those can be excited at high temperature.
For the 4d description to be valid, we should have the
temperature smaller than $M_c$. This requirement 
is in addition to the 5d cut-off $\Lambda$ which
characterizes the scale at which the 5d gauge theory (including KK modes) becomes
strongly coupled, and loses predictivity.

The definition of the frontier between the 4d and 5d descriptions in
the inflationary phase, where the universe is dominated by vacuum
energy, is somewhat subtle.  During inflation it is possible to define
a temperature. In the exponentially expanding universe particles are
created with typical momentum of the order of $H$. In this phase, the
criterion for the validity of the 4d effective description is
$H\lesssim L^{-1}_{phys}$, where $L_{phys}=e^{\langle \phi \rangle
/\sqrt{3}}/M_c$.  This leads to a constraint on $M_c$, shown in Fig.\
\ref{fig:5dbound}. The corresponding constraint is weak, and is only
relevant for small values of r, which, as we will see in the next
section, will be disfavored anyway because of the problem of the
overclosure of the universe by radions.

The reheat temperature satisfies $T_{RH}\lesssim M_I$ 
where the bound is saturated in the case of preheating.  
We can define a parameter ${\cal V}\equiv T_{RH}/M_I$ 
which measures the efficiency of the transfer between the inflaton 
energy to radiation energy.  Requiring an effective 4d description 
($T_{RH}\lesssim M_c$) leads to ${\cal V}\lesssim r$.
Note that in the case of perturbative reheating 
$T_{RH}\sim \lambda \sqrt{m_I m_{Pl}}$ where $\lambda$ is the inflaton 
coupling to normal matter and $m_I$ is the inflaton mass. 
The requirement that the theory appear four dimensional 
thus translates into $m_I\lesssim r^2\lambda^{-2} 10^{11}$ GeV.

\EPSFIGURE[t]{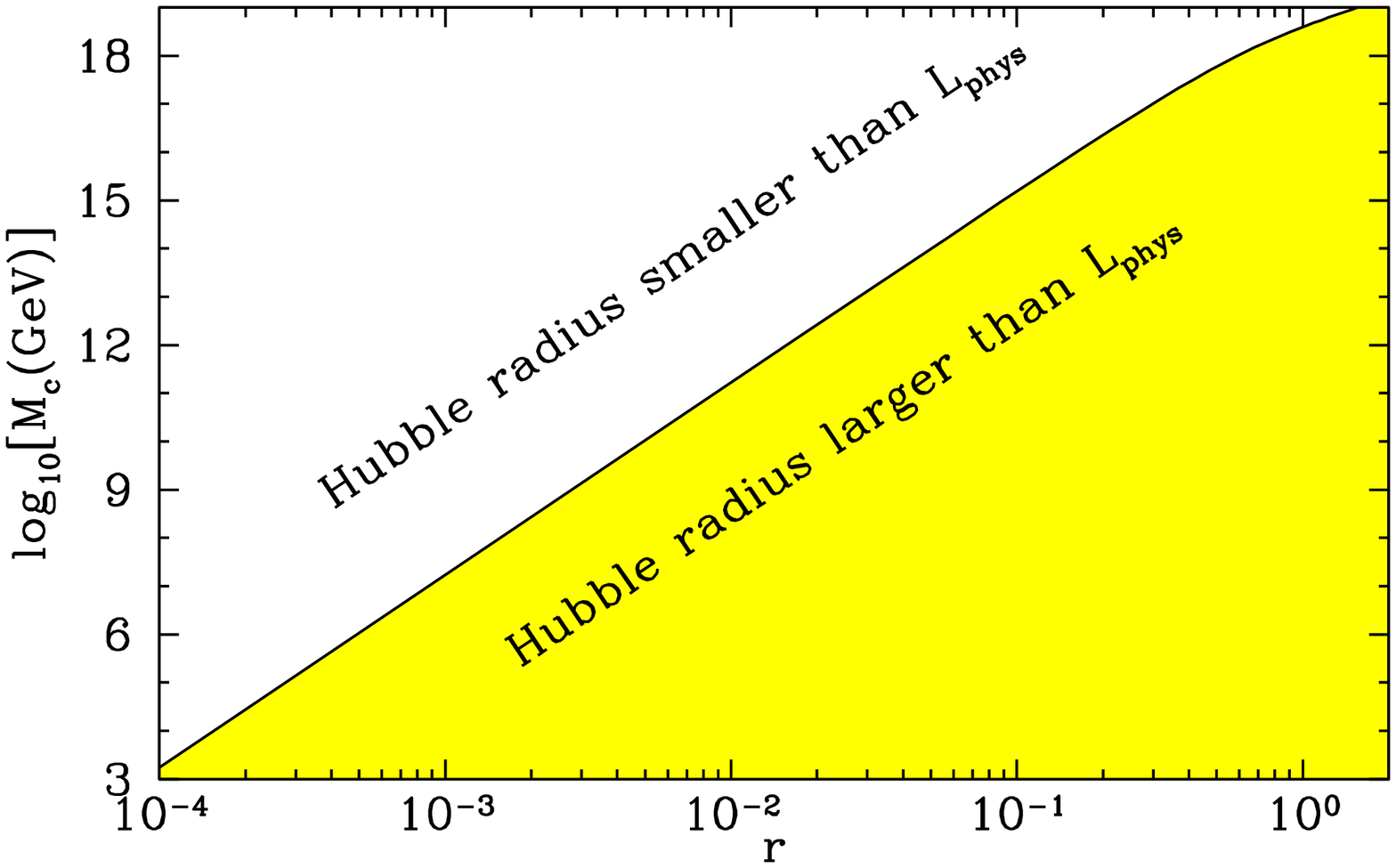,width=15cm}
{For the Hubble radius to be larger than the size of the extra dimension 
$L_{phys}=e^{\phi/\sqrt{3}}/M_c$ during inflation, $M_c$ must 
be in the lower region limited by the dashed line. This does not lead 
to strong constraints in the preferred region $r\gtrsim 1$.
\label{fig:5dbound}}

\section{Radion as Dark Matter}
\label{sec:radionDM}

\subsection{Coherent Production}

We now turn to the evolution of the radion field once inflation is
completed. The story begins with the energy density of the radion
after inflation (determined by the parameter $D$ discussed in the
previous section).  From the time oscillations commence until 
today,\footnote{If the radion is dark matter it must have a
lifetime longer than the age of the universe, which requires
$M_c\lesssim 7 \times 10^8$ GeV.} the radion energy density $\rho_{r}$
redshifts as nonrelativistic matter, so that
\begin{equation}
\label{e1}
\rho_{r}(t_0)=\rho_r(t_{osc})\frac{a_{osc}^3}{a_0^3} ,
\end{equation}
where the subscript `0' refers to the present, and
\begin{equation}
\rho_r(t_{osc})=\frac{1}{2} m_r^2\phi^2(t_{osc})= \frac{3}{2}D^2 M_c^4.
\end{equation}

After inflation, until reheating (or preheating), the energy densities
in both the inflaton and the radion decrease as a nonrelativistic
component.  We can express $\rho_r(t_0)$ as
\begin{equation}
\rho_r(t_0) = \frac{3}{2}D^2 M_c^4 \frac{a_{osc}^3}{a_{RH}^3}
				   \frac{a_{RH}^3}{a_0^3}.
\end{equation}
Before reheating, $a^3\propto t^2$. Since $t^2\sim H^{-2}\propto
\rho^{-1}$, we can replace the factor $a^3_{osc}/a^3_{RH}$ by
$\rho_{RH}/\rho_{osc}$.  The value of $\rho_{RH}$ is simply
$(\pi^2g_*(T_{RH})/30)T_{RH}^4$.  The value of $\rho_{osc}$ is not so
straightforward.  

Here we will use the results for the chaotic inflation example of the
previous section.  If $r \lesssim 1$, then the radion is the inflaton.
In this case inflation ends when $\phi \simeq m_{Pl}$ and
$\rho_{osc}\simeq m_r^2m_{Pl}^2$.  Also in this case, $D\simeq
10^{-1}$.  If $r \gtrsim 1$ the energy during inflation is
approximately $M_I^4$ and $D\simeq 10^{-2}r^{-4}$ (as can be seen from
Fig.\ \ref{fig:Dofr}). Using these values, along with the fact that
$a_{RH}^3/a_0^3=g_*(T_0)T_0^3/g_*(T_{RH})T_{RH}^3$, we find
\begin{equation}
\rho_r(t_0)=\frac{\pi^2}{20}g_*(T_0)T_0^3T_{RH}\left\{ \begin{array}{ll}
	3 \times 10^{-3} & \quad r \lesssim 1 \\
	10^{-4}r^{-4} & \quad r \gtrsim 1 .
        \end{array}\right.
\end{equation}
Using $T_0=2.4 \times 10^{-13}$ GeV, $g(T_0)=3.36$, and
$\rho_c=3.9\times 10^{-47}$ GeV$^4$, this leads to a value of
$\Omega_r$ of
\begin{equation}
\Omega_r = \left\{ \begin{array}{ll}
     2\times 10^6\ (T_{RH}/\textrm{GeV}) & \quad r\lesssim 1\\
     6\times 10^4\ (T_{RH}/\textrm{GeV})\ r^{-4}& \quad r\gtrsim 1.\\
	\end{array} \right.
\label{eq:Omegaradion}
\end{equation}

If the radion is stable, we require $\Omega_r
\lesssim 0.3$, and the inequality is saturated if the radion is dark matter.
The limit to $\Omega_r$ results in the limit $T_{RH} \leq
5\times10^{-7}$GeV for $r \lesssim 1$, and $T_{RH}\leq
2\times10^{-5}r^4$GeV for $r \gtrsim 1$.  Clearly the $r\lesssim 1$
case is ruled out.  The parameter $r$ must be sufficiently large to
avoid overclosure:
\begin{equation}
r \gtrsim 15 \left(\frac{T_{RH}}{\textrm{GeV}}\right)^{1/4}.
\label{Thebound}
\end{equation}

We now discuss some potential observational effects associated with the
oscillation of the radion\footnote{See 
also \cite{Perivolaropoulos:2002pn,Perivolaropoulos:2003we}}.  
Coherent oscillations of the radion induce
variations of gauge couplings (in a way described in Ref.\
\cite{Chacko:2002sb}, for instance). For this effect not be observable, we
require the oscillation period $m_r^{-1}$ to be much smaller than the
Hubble time. There are constraints on the variation of the gauge
couplings at the time of nucleosynthesis. Applying the above
requirement to the nucleosynthesis epoch leads to $H_{nucleo}\ll m_r$,
{\em i.e.,} $T_{nucleo}\ll M_c$, which is always satisfied.  Note that
as soon as structures start to form, coherent oscillations will get
distracted by the development of gravitational potentials. Therefore,
radion oscillations would no longer be coherent today and would not be
seen as variations of the couplings, but instead, as nonrelativistic
matter.

\subsection{Thermal Production}

Now we turn to the thermal production of radions at reheating.
The number density of radions produced thermally is obtained by
solving the Boltzmann equation
\begin{equation}
\label{eqn0}
\frac{dn_r}{dt}+3Hn_r=-\langle \sigma v\rangle n_\gamma 
(n_r-n_r^{eq})-\Gamma(n_r-n_r^{eq}),
\end{equation}
where $n_\gamma$ represents the number density of a typical light
species in the reaction $\mbox{radion}+ \gamma \rightarrow$
everything.  The cross section is expected to be approximately
$m_{Pl}^{-2}$, and the decay width is of order
$m_r^3/m_{Pl}^2$.\footnote{There are potentially large dimensionless
numbers in the cross section and decay width; {\em e.g.,} the factor
of $192\pi$ appearing in the decay width given in Eq.\ (\ref{lifetime}).}
The feeble radion interactions with matter imply that the radion has
 great difficulty reaching thermal equilibrium. We have also seen above 
 that the coherent production of radions
 need be very small. Thus, initially, $n\ll n_{eq}$. 
In terms of $Y\equiv n_r/s$ where $s$ is the entropy density, and
$x\equiv m_r/T$, the Boltzmann equation becomes
\begin{equation}
\label{eqn9}
\frac{dY}{dx} = \frac{m_r}{m_{Pl}}\left( \frac{C_A}{x^2}+xC_D\right)Y_{eq},
\end{equation}
where $C_A$ and $C_D$ are dimensionless numbers.

If the radions are relativistic ($x\ll1$), $Y_{eq}\sim x^{-3}$, and 
\begin{equation}
Y \sim \frac{T_{RH}}{m_{Pl}}.
\end{equation}
We note again that there is a potentially small (say $10^{-2}$)
dimensionless number on the right-hand-side of the above equation.

We now examine the case where the radions are nonrelativistic at the
time of reheating ($x\gg1$) so that $Y_{eq}\sim x^{3/2}\exp(-x)$ and
we can ignore the first term in the right hand side of Eq.\
(\ref{eqn9}).  In this case
\begin{equation}
\label{non_relat}
Y\sim \frac{m_r}{m_{Pl}} \ {\cal I}(x_{RH}),
\end{equation}
with
\begin{equation}
{\cal I}(x_{RH})\equiv \int_{x_{RH}}^{\infty}x^{5/2}e^{-x}dx,
\end{equation}
where $x_{RH} = m_r / T_{RH}$ is $x$ at the time of reheating.

We find that $\Omega_r$ from thermal radions (for relativistic radion
production) is
\begin{equation}
\Omega_r= 10^9\frac{T_{RH}}{\mbox{GeV}} \left(\frac{M_c}{m_{Pl}} \right)^2,
\end{equation}
while the value for nonrelativistic radion production is 
\begin{equation}
\label{thermalnr0}
\Omega_r=10^9\mbox{GeV}^{-1}\frac{M_c^4}{m_{Pl}^3} \times {\cal I}(x_{RH}).
\end{equation}

The limit $\Omega_r<0.3$ becomes
\begin{equation}
\left( \frac{M_c}{5\times10^{13}\mbox{GeV}} \right)^2 
\frac{T_{RH}}{\mbox{GeV}} < 1
\end{equation}
for relativistic radion production, and for nonrelativistic radion production
\begin{equation}
\label{thermalnr}
\left( \frac{M_c}{4 \times 10^{11}\mbox{GeV}} \right)^4 {\cal I}(x_{RH})< 1 .
\end{equation}
This last bound is always satisfied, while the bound from relativistic radions 
only leads to a very weak constraint on the reheat temperature. 
Therefore, there is no consequence, as far as dark matter is concerned, 
from thermally produced radions. 

Radionactive dark matter would be challenging to detect directly,
though it may have implications either for gamma-ray spectra (through
the eventual decay of the radions) or isocurvature perturbations.

\section{Radions and Cosmological Perturbations}
\label{sec:curvaton}

\subsection{Radion Isocurvature Perturbations}

A possible effect from the stable radion is on the CMB, an issue we
now investigate. Indeed, in a way reminiscent of the axion, there are
isocurvature perturbations associated with the radion as dark
matter. To discuss this, we begin with a few definitions.  The
curvature perturbation associated with an energy density component
$\rho_i$ is defined by
\begin{equation}
\zeta_i=-H\frac{\delta\rho_i}{\dot{\rho}_i} .
\label{def_zeta_i}
\end{equation}
For matter, $\zeta_m=\frac{1}{3}\delta\rho_m/\rho_m$ and for radiation
$\zeta_{\gamma}=\frac{1}{4}\delta\rho_{\gamma}/\rho_{\gamma}$. The
adiabatic mode of the perturbation is defined by
$\zeta_i=\zeta_{\gamma}$. The nonadiabatic or isocurvature modes
arise from local inhomogeneities in the equation of state of the
energy density.  They may be specified relative to $\zeta_{\gamma}$ as
\begin{equation}
{\cal S}_i\equiv\frac{\delta \rho_i}{\rho_i}-\frac{3}{4}
\frac{\delta \rho_{\gamma}}{\rho_{\gamma}}
=3(\zeta_i-\zeta_{\gamma})
\end{equation}
The large-scale temperature anisotropy can be related to the
primordial adiabatic and isocurvature perturbations by (see for
instance section II of Ref.\ \cite{Gordon:2002gv}):
\begin{equation}
\frac{\delta T}{T}\approx-\frac{1}{5}\zeta^{ad}-\frac{2}{5}
\left(\frac{\rho_{c}}{\rho_m}{\cal S}_c
+\frac{\rho_b}{\rho_m}{\cal S}_b\right) +\frac{1}{15}
\frac{\rho_{\nu}}{\rho_{\gamma}+\rho_{\nu}}{\cal S}_{\nu}
\end{equation}
where $\zeta^{ad}$ measures the primordial adiabatic perturbation, 
${\cal S}_c$, ${\cal S}_b$ and ${\cal S}_{\nu}$ the isocurvature 
perturbations in 
cold dark matter, baryons and neutrinos respectively. Here
$\rho_m$ is the total matter energy density, $\rho_m=\rho_c+\rho_b$.

We are interested in constraining ${\cal S}_c$. From an 
observational point of view, what is constrained is the quantity:
\begin{equation}
B=\frac{{\cal S}_b^{\rm{eff}}}{\zeta^{ad}},
\end{equation}
where
\begin{equation}
{\cal S}_b^{\rm{eff}}={\cal S}_b+\frac{\rho_c}{\rho_b}{\cal S}_c
\end{equation}
is the effective nonrelativistic term.  We use the constraints
derived at 95\% C.L.\ in Ref.\ \cite{Gordon:2002gv}
\begin{equation}
-0.46 < B < 0.35  .
\label{isocons}
\end{equation}
In the case where the radion is the dark matter we have 
\begin{equation}
{\cal S}_c={\cal S}_{radion}=3(\zeta_{radion}-\zeta_{\gamma})
=\frac{\delta\rho_{\phi}}{\rho_{\phi}}=\delta^{iso},
\end{equation}
 where, at first order,
\begin{equation}
\label{deltaiso}
\delta^{iso}=2\frac{\delta \phi}{\phi} .
\end{equation}
If the radion is practically free during inflation, with small
 effective mass $m^2\ll H^2$, it follows that on superhorizon scales
\begin{equation}
\frac{\delta \phi}{\phi}\approx\frac{H_*}{2 \pi \phi_*},
\end{equation}
where $H_*$ and $\phi_*$ are $H$ and $\phi$ evaluated at the time of horizon
exit,  and
\begin{equation}
\delta^{iso}=\frac{H_*}{\pi \phi_*}=\frac{\sqrt{V_*}}
{\sqrt{3}\pi D_* m^2_{Pl}},
\end{equation}
where $D_*$ is $\phi_*/m_{Pl}$.  Equation (\ref{isocons}) leads to the
restriction $|\delta^{iso}|\lesssim10^{-2}\zeta_{\gamma}$, which leads to
\begin{equation}
\label{isocons2}
\sqrt{V_*}\lesssim 10^{-5} D_* m_{Pl}^2.
\end{equation}
This constraint, while interesting in its own right, is not relevant
for our current picture.  As we will see in the next section,
$m^2/H^2 \lesssim 1$ during inflation only if $r\lesssim 0.4$, which
typically results in $D$ of order unity. But if $D$ is of order unity,
the limit in Eq.\ (\ref{isocons2}) is weaker than the limit on $V_*$
from the limit to the gravitational wave background produced in
inflation.  However, we note that there may exist unconventional
inflationary mechanisms where this limit is useful.

To summarize, we have seen that when the compactification
scale is below $7 \times 10^8$ GeV and the radion is effectively
stable, to avoid overclosure of the universe by radions we must have
$r \gtrsim 15(T_{RH}/\textrm{GeV})^{1/4}$.

If $D$ and $r$ are properly correlated, radions could in fact play
the role of cold dark matter, though in the models of inflation we
have considered this would require fine tuning.

\subsection{Radion as the Curvaton}

It is now widely believed that the origin of structure in the universe
is a primordial perturbation already in existence when cosmological
scales start to enter the horizon.  In the standard picture for the
generation of cosmological perturbations, the curvature perturbation
$\zeta$ is generated during inflation through the perturbation of a
single component inflaton. 

In the curvaton scenario \cite{Lyth:2001nq} the curvature perturbation
produced during inflation is initially negligible, but the curvaton
field acquires ``isocurvature'' perturbations during inflation, which
is converted to curvature perturbations when the curvaton density
becomes a significant fraction of the total density.  In this case,
$\zeta$ is given by
\begin{equation}
\zeta=f_D \times \frac{\delta^{iso}}{3}
\end{equation}
where $f_D$ is the fraction of the total energy density (radion $r$ and
radiation $R$) of the universe in the radion--curvaton when it decays:
\begin{equation}\label{f}
f_D= \left[ \frac{\rho_r}{\rho_r+\frac{4}{3}\rho_R}\right]_{H=\Gamma} .
\end{equation}

When the radion decays, it reheats the universe and imprints its
perturbation into photons and matter. The degree of nonadiabaticity is
given by the fraction $f_D$ which reflects how much of the total energy
density is carried by the radion at the time of its decay. If this
fraction $f_D$ is unity, there will be no residual isocurvature
perturbations. On the other hand, if the radion does not completely
dominate the energy density of the universe at the time of decay,
there will be residual isocurvature perturbations.

If $m^2\ll H^2$ during inflation then
the spectrum of the fractional field perturbation is
\begin{equation}
{\cal P}^{1/2}_{\delta \phi / \phi }\approx \frac{H_*}{2 \pi \phi_*} 
\end{equation}
and that of the fractional energy density perturbation is
\begin{equation}
{\cal P}^{1/2}_{\delta^{iso}}\approx \frac{H_*}{ \pi \phi_*} 
\approx \delta^{iso} ~~ .
\end{equation}
It follows that the prediction of the radion-curvaton model for the 
spectrum of the curvature perturbation is
\begin{equation}
{\cal P}^{1/2}_{\zeta}\approx \frac{f_D}{3} \delta^{iso}
\approx\frac{f_D H_*}{3 \pi \phi_*}.
\label{zeta_final}
\end{equation}
This is to be compared with the COBE measurement of the CMB quadrupole
anisotropy
\begin{equation}
{\cal P}^{1/2}_{\zeta}=4.8 \times 10^{-5} ~~ .
\end{equation}

It is now clear that the radion cannot easily play the role of a
curvaton. Indeed, from Fig.~\ref{fig:ratio}, 
we see that the effective mass of the radion is smaller than $H$ only 
for $r\lesssim 0.4$. On the other hand, Fig.~\ref{fig:potentials}
indicates that for small $r$, $\phi$ dominates the total energy 
density during inflation, in conflict with the curvaton scenario.
In fact, this leads back to the model of the radion as the inflaton 
(and providing the primordial fluctuations) discussed in 
Sec.~\ref{subsec:radinflaton}.

It is possible that $\psi$ could play the role of the curvaton. This
would require that $\psi$ decays after $\phi$, which may be unlikely
for low compactifications scales since generically a scalar field
would have gravitational interactions, and thus at most a lifetime of
the same order as the radion.  This would require $\lambda
m_{\psi}\lesssim M_c^6/m_{Pl}^5$, where $\lambda$ characterizes the
$\psi$ interaction strength.

Let us now see if the radion can play the role of a {\it modulon}.

\EPSFIGURE[t]{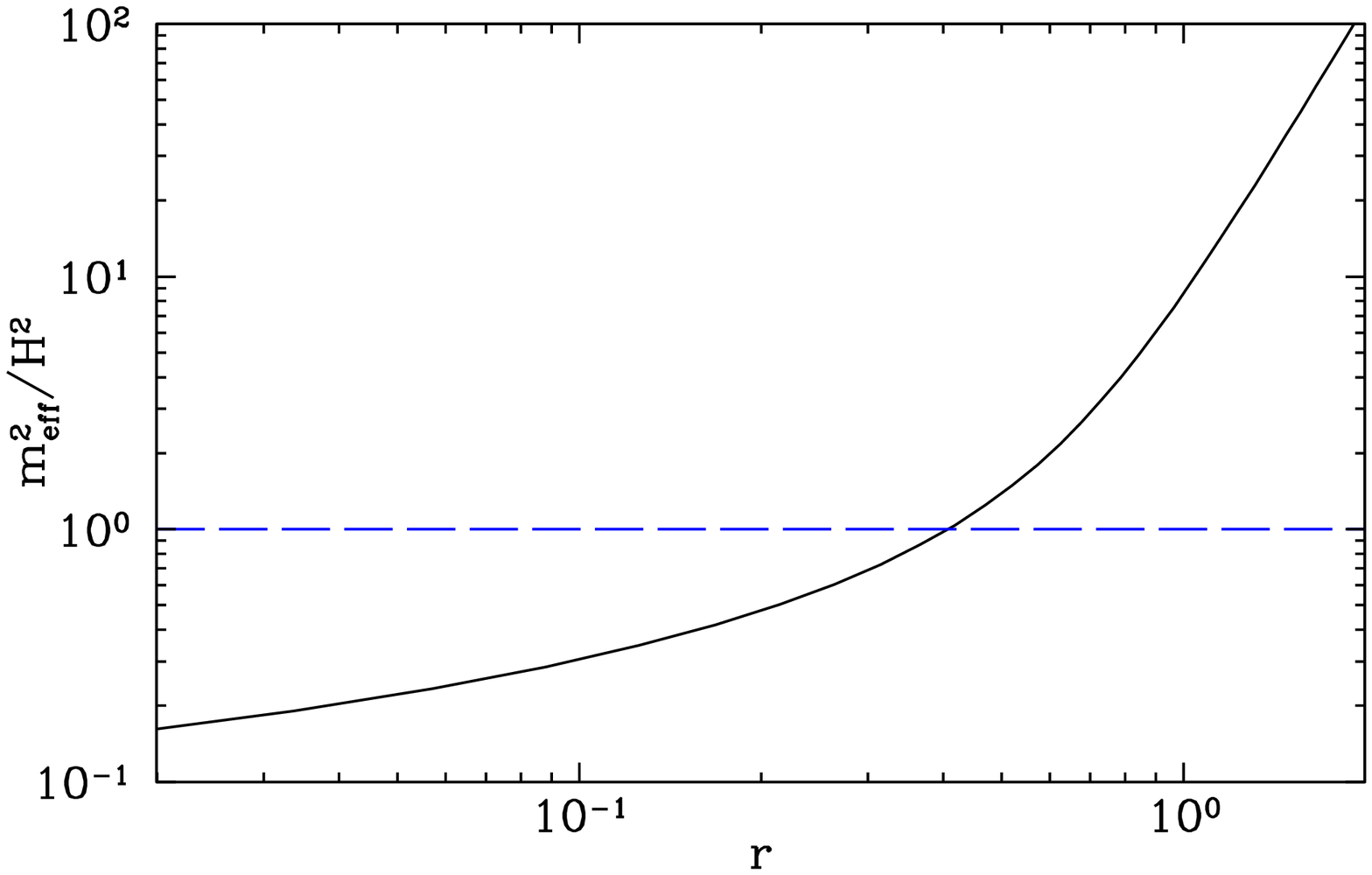,width=15cm}
{The ratio $m_{\rm{eff}}^2/H^2$ during inflation (evaluated at the
high-energy minimum of the potential for $\phi$). Fluctuations in
$\phi$ are frozen and can be at the origin of cosmological
perturbations only for $r\lesssim 0.4$. \label{fig:ratio}}

\subsection{Radion as the Modulon}
\label{sec:modulon}

As mentioned in the introduction, another alternative to the standard picture
of generation of cosmological fluctuations was recently proposed
\cite{Dvali:2003em}, which also makes use of a light scalar
field that controls the coupling of the inflaton to matter.  
To distinguish it from the curvaton scenario, we refer to the field
responsible for the generation of fluctuations in this mechanism as a {\it
modulon}. In the modulon scenario, the coupling $\lambda$ between the inflaton
and ordinary matter is not constant but is given by the fluctuations around the
VEV of the modulon.  If the modulon is light enough, it acquires primordial
fluctuations during inflation of order $H_*/2\pi$. These spatial fluctuations
will then correspond to spatial fluctations in the coupling $\lambda$. In the
following we will denote the modulon by $\phi$ since we want to check whether
the radion can be a modulon candidate.  The (local) reheat temperature 
depends on this coupling according to
\begin{equation}
T_{RH} \sim \sqrt{\Gamma_I m_{Pl}} ,
\end{equation}
where $\Gamma_I\sim \lambda^2 m_I$ is the decay rate of the inflaton and $m_I$
its mass.  In the modulon scenario,
\begin{equation}
\frac{\delta T}{T}\sim \frac {\delta \lambda}{\lambda} ,
\label{modfluc}
\end{equation}
where $\lambda$ is given by
\begin{equation}
\lambda(\phi)=\lambda_0(1+\frac{\phi}{M}+ ...) .
\label{moduloncoupling}
\end{equation}
As the radion controls the size of the extra dimension, it
controls the inflaton coupling to matter as in Eq.~(\ref{moduloncoupling}),
where $\lambda_0$ is related to a dimensionful, 5d parameter by 
$\lambda_0 = \lambda_5 / L$.  Small variations in the radion VEV thus
indeed lead to variations in the coupling between matter and the
inflaton, with $M=\sqrt{3}m_{Pl}$.
The fraction of the coupling controlled by the fluctuating radion is
$\phi_*/{m_{Pl}}$ where $\phi_*=D_* m_{Pl}$ is the value of $\phi$ at the time
of horizon exit so that Eq.~(\ref{modfluc}) can be written as
\begin{equation}
\frac{\delta T}{T}\sim \frac{\phi_*}{m_{Pl}}\frac{H_*}{2\pi \phi_*}
\sim \frac{H_*}{2\pi m_{Pl}} ,
\end{equation}
a contribution typically much too small to account for observed 
inhomogeneities.  Thus, in models with flat extra dimensions, the radion 
cannot effectively play the role of a modulon. For this mechanism to be
effective, the radion coupling should be much larger, as for example in warped
compactifications. In the Randall-Sundrum geometry, we would arrive at
${\delta T}/{T}\sim {H_*}/{ 1 \ \mbox{TeV}}$. In order
to explain the COBE result, this would require an inflation scale around 
$10^8$ GeV, which is neither the fundamental UV nor the IR scale of the 
Randall-Sundrum model, but could be generated by an intermediate brane
\cite{Collins:2002kp}.

\section{Late Decay of Radions and KK Gravitons}
\label{sectionBBN}

The decay of massive relics after Big Bang Nucleosynthesis produces
energetic photons which may dissociate light elements 
and modify the predictions of their abundances.
The constraint on the decay of heavy relics
from BBN is (see for instance \cite{Ellis:1990nb})
\begin{equation}
\label{BBNcons}
mY\lesssim 10^{-12} \ \ \mbox{GeV}
\end{equation}
where $Y\equiv n_X/s$ is the ratio we would have observed today 
if particle {\em had not decayed,} and $s\sim T^3$ is the entropy per
comoving volume.  This bound literally applies for particles whose
lifetime is $10^8$ s; bounds for larger lifetimes are somewhat relaxed.
Thus, by taking $10^{-12}$ GeV on the right hand side of Eq.~(\ref{BBNcons})
we are making a conservative estimate for the allowed parameter space.
Note that the constraint Eq.~(\ref{BBNcons}) applies only if the decay 
takes place after nucleosynthesis.

\subsection{Late Decay of Radions}

{}From our estimate of the radion lifetime, Eq.~(\ref{lifetime}), we find that
the radion decays after nucleosynthesis ($\tau \gtrsim 10^{-2}$;
$\Gamma \lesssim 10^{-22} \ \mbox{GeV}$) provided
\begin{equation}
7 \times 10^{8} \mbox{GeV} \lesssim M_c \lesssim  10^{12} \mbox{GeV} ,
\label{bbnbound}
\end{equation}
where the lower limit on $M_c$ comes from the requirement that the
radion lifetime is smaller than the age of the universe. In order to
explore the implications on radion cosmology , we now estimate
$Y_r=n_r/s$ at the time of reheating (which is the same as the ratio we
would have observed today if the expansion is adiabatic).

The bound in Eq.~(\ref{BBNcons}) may be written as
\begin{equation}
m_r \frac{n_r}{s}=\frac{\rho_r}{s}=
\Omega_r \frac{\rho_c}{s} \leq 10^{-12} \mbox{GeV} .
\end{equation}
Using $\rho_c/s\sim 10^{-9}$, we obtain the constraint $\Omega_r
\lesssim 10^{-4}$.  

For the chaotic inflationary potential, coherent
radion production [using Eq.\ (\ref{eq:Omegaradion})] yields
\begin{eqnarray}
\frac{T_{RH}}{\mbox{GeV}} \lesssim \left\{ 
\begin{array}{ll}
5\times10^{-11} & \quad r\lesssim 1\\
2\times10^{-9} r^4 & \quad r\gtrsim 1.\\
\end{array} \right.
\label{nonthermal}
\end{eqnarray}
Recall that this bound only applies in the range $7 \times 10^8 \ \mbox{GeV}
\lesssim M_c\lesssim 10^{12} \  \mbox{GeV}$, so that the radion decays after
nucleosynthesis.   Obviously the case $r\lesssim 1$ is again excluded. The bound on 
$T_{RH}$ for $r\gtrsim 1$ is plotted on figure \ref{fig:DMconstraintonD2}.

The constraint in Eq.\ (\ref{BBNcons}) also applies to radions of
thermal origin:
\begin{equation}
T_{RH}\lesssim \left(\frac{10^{12}\mbox{GeV}}{M_c}\right)^2 \mbox{GeV} 
\end{equation}
from the thermal production of relativistic radions in the radiation era.
If $M_c\sim 10^9$ GeV, then $T_{RH}\lesssim 10^6$ GeV. If $M_c\sim
10^{12}$GeV, then the bound becomes $T_{RH}\lesssim 1$ GeV.  

For nonrelativistic radion production [using Eq.\ (\ref{thermalnr0})]
the bound in Eq.\ (\ref{BBNcons}) becomes
\begin{equation}
{\cal I}(x_{RH})
\lesssim \left(\frac{4 \times 10^{10} \: \mbox{GeV}}{M_c}\right)^4 .
\end{equation}
This bound imposes constraints on $T_{RH}$ for the larger region of
$M_c$.  Below $M_c \lesssim 3 \times 10^{10}$ GeV, there is no bound
on $T_{RH}$.  At $M_c \sim 10^{12}$ GeV, the bound is $T_{RH} \lesssim
35$ TeV, for $M_c \sim 10^{11}$ GeV, the bound is $T_{RH} \lesssim
800$ GeV and for $M_c \sim 5 \times 10^{10}$ GeV, it is $T_{RH} \lesssim
320$ GeV.

\EPSFIGURE[t]{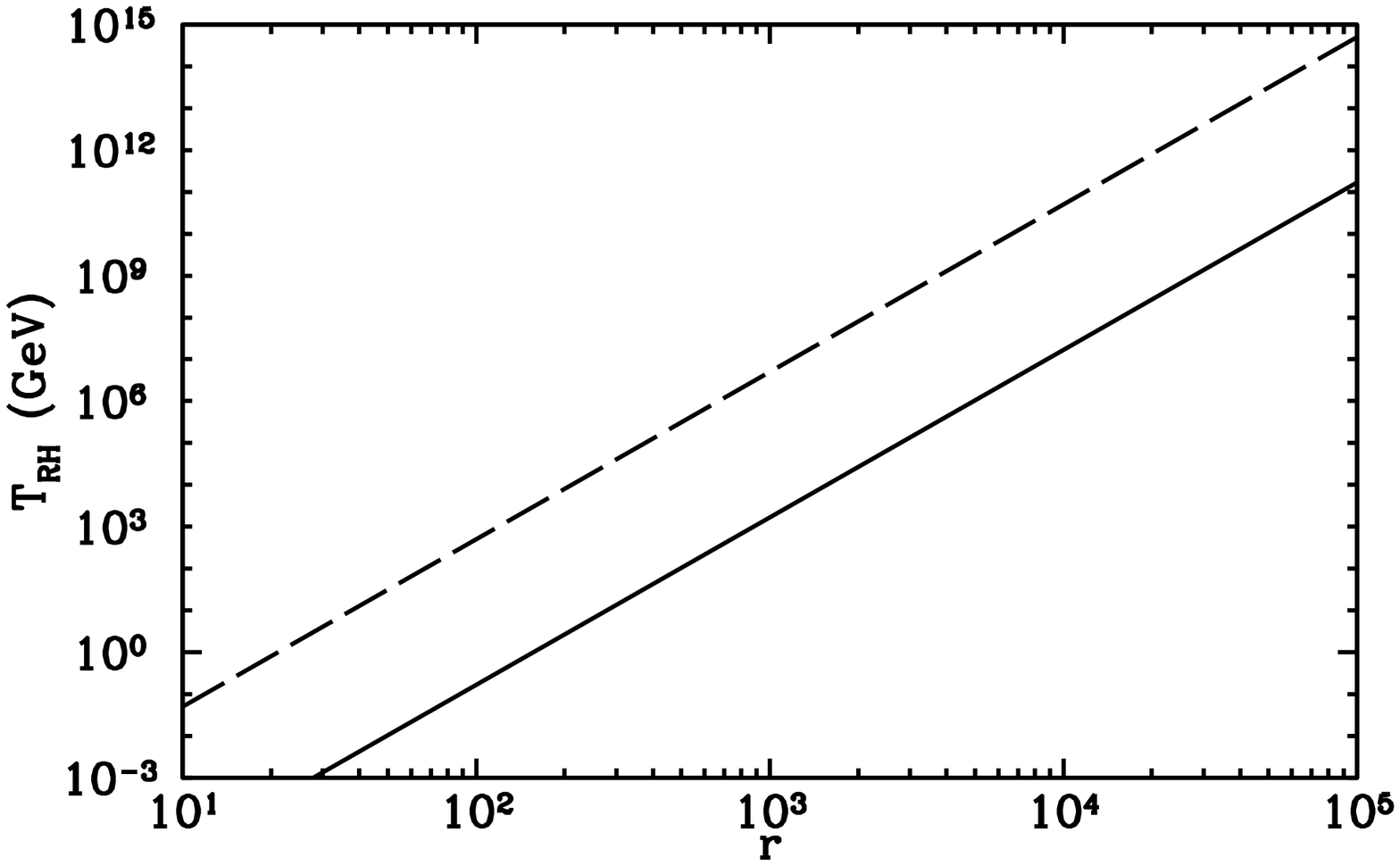,width=15cm}
{The constraints from overclosure by radions (produced 
non-thermally) --dashed curve-- as well as
from BBN --solid curve-- due to the late decay of radions, translated
into a bound on $T_{RH}$ as a function of $r$.  The bound from
overclosure (see Eq.~\ref{eq:Omegaradion}) applies in the range $M_c \sim 
10^3-7 \times 10^8$ GeV and
the one from BBN (Eq.~\ref{nonthermal}) in the range $M_c \sim 7 \times 10^8-10^{12}$ GeV.
\label{fig:DMconstraintonD2} }

\subsection{Late Decay of KK gravitons}
\label{sub:KKgravitons}

KK gravitons ($G^1$) also couple very weakly, $\sim m_{Pl}^{-1}$. 
They have a mass characterized by the size of the extra dimension,
$m_{KK}\sim M_c$ and therefore a decay width of the order 
$\Gamma \sim M^3_c/m^2_{Pl}$, though the specific details of the expression
depend on whether KK parity is well conserved, which would forbid 
decays directly into zero modes of SM fields.  Continuing with dimensional
analysis, the decay is after nucleosynthesis if $M_c\lesssim 8 \times
10^4$ GeV.  On the other hand, we only consider values of $M_c$ above several
hundred GeV, as lower values are generally excluded by collider constraints
\cite{Appelquist:2002wb}.
Thus, constraints on KK gravitons from BBN will only apply in the narrow range
\begin{equation}
300 \ \mbox{GeV} \lesssim M_c\lesssim 8 \times 10^4 \ \mbox{GeV}
\end{equation}
In this range of values for $M_c$, the radion is stable, and to 
avoid overclosure of the universe by radions, $r$ ($D$) and $T_{RH}$
must be chosen appropriately in Eq.~(\ref{eq:Omegaradion}).

We assume the universe is 
radiation dominated at reheating, and compute bounds from the thermal 
production of KK gravitons.  As for the radion, 
$\langle \sigma v \rangle \sim m_{Pl}^{-2}$, and many of the radion
results can be carried over with $m_r \rightarrow M_c$.
We begin with the case of relativistic KK gravitons,
$T_{RH} \gtrsim M_c$.  In this case,
thermal production of KK gravitons leads to 
$Y\sim T_{RH}/m_{Pl}$ and the BBN bound results in,
\begin{eqnarray}
T_{RH} & \lesssim & \left( \frac{10^{6} \mbox{GeV}}{M_c} \right) \mbox{GeV} 
\end{eqnarray}
from the late decay of thermally produced, relativistic KK gravitons.  For
$M_c$ at the TeV scale, this roughly constrains $T_{RH}$ to be of order
$M_c$.  For somewhat smaller extra dimensions, this requires a smaller reheat
temperature, and indicates that KK gravitons cannot in fact be relativistic for
$M_c \gtrsim 1$ TeV.

For nonrelativistic KK gravitons, $Y \sim M_c / m_{Pl} x_{RH}{\cal I}$,
where now $x_{RH} = M_c / T_{RH}$.  Thus, the bound on the reheat temperature
is,
\begin{eqnarray}
{\cal I} \ \frac{M_c}{T_{RH}}  \lesssim 
\left( \frac{10^3 \mbox{GeV}}{M_c} \right)^2
\end{eqnarray}
which for all $M_c$ of interest results in a bound of 
$T_{RH} \lesssim M_c /{\rm a \; few} $\footnote{This point was realized some time ago by
Abel and Sarkar \cite{Abel:1994bv}},
consistent with the possibility of KK dark matter \cite{Servant:2002aq}, 
and in fact with the general WIMP hypothesis, for which the freeze-out
temperature is typically a few tens of GeV. 
Note also that Ref.~\cite{Feng:2003xh} studied the case where KK dark matter 
was made of KK gravitons rather than KK hypercharge gauge bosons, with the
relic abundance determined initially by freeze-out of $B^1$. 
They derived bounds from BBN on this scenario coming from the eventual 
late decay of $B^1$ into $G^1$ and a photon.

To finish this section, we discuss the possibility of nonthermal 
production of KK gravitons. As stressed in 
\cite{Giudice:1999yt}, the gravitational
production of particles during inflation can be very efficient and can
lead to much stronger bounds on the reheat temperature than the ones
from thermal production. However, for this mechanism to be
operational, we need the mass of the particle to be smaller than $H$
during inflation. In our case, this corresponds to $e^{-\phi/3}M_c<H$.

{}From Fig\ \ref{fig:5dbound}, this leads to a bound on $M_c$ only for
very small values of $r$. On the other hand, because of the bound from
overclosure of the universe we typically have $r\gtrsim 1$ so that no
KK gravitons are expected to be produced nonthermally. Note also that
from Fig.\ \ref{fig:ratio}, there is essentially no gravitational
production of radions during inflation for $r\gtrsim 1$. In any case,
the bound from overclosure of the universe by radions is not the only
reason why a small hierarchy between $M_c$ and $M_I$ is required.  A
large enough reheat temperature, to have WIMP dark matter, also
requires large values of $r$.
 
\subsection{Diffuse Gamma-ray Signal}

Finally, for a very long lived radion, there might be some interesting
observable diffuse gamma-ray signal \cite{Ellis:1990nb,Kribs:1996ac}. This
applies for life times larger than $10^{14}$s, {\em i.e.,} for
\begin{equation}
M_c\lesssim 3\times 10^9 \ \mbox {GeV}.
\end{equation}
On the other hand, the radion becomes stable for $\tau\gtrsim 10^{17}$s, {\em
i.e.,} for $M_c\lesssim 9 \times 10^8$ GeV. The observable signal therefore
corresponds to a tiny window of values for $M_c$. This is because the decay
rate depends on the sixth power of $M_c$. Note that there is no similar
expected signal from decaying KK gravitons 
(however see \cite{Feng:2003xh} for the special case of UED)
since $\tau \gtrsim 10^{17}$s
corresponds to the disallowed region $M_c\lesssim0.4$ GeV.

\section{Conclusions}
\label{sec:conclusion}

In this work we have studied the cosmological properties of the
radion field as it appears in models with flat or almost flat extra
dimensions.  We find that the potential roles the radion may play in
cosmology are very sensitive to the scale (and model) of inflation.
Thus, our work explores the challenges involved in building a
realistic cosmological model for theories with flat extra dimensions.
Figure \ref{fig:summary} summarizes our results.

Its origin as a component of the higher dimensional graviton tensor
dictates the radion's couplings.  Similarly, the mass of the radion, in
the absence of fine tuning, is generally related to the
compactification scale $M_c$ by $m_r \sim M_c^2 / m_{Pl}$.  Thus, once
one has chosen the size of the extra dimension, the natural radion
properties are determined.  Of course, it is possible that the radion
potential is itself fine tuned by some mechanism we currently do not
understand.  In that case, its couplings to other fields may still be
of gravitational strength, but its mass could possibly differ from our
estimates.  One avenue of future research would be to explore the
compactification dynamics that could naturally make the radion, for
example, more massive than the na\"{\i}ve estimates, such that it
decays at early times and evades some cosmological constraints.

\EPSFIGURE[t]{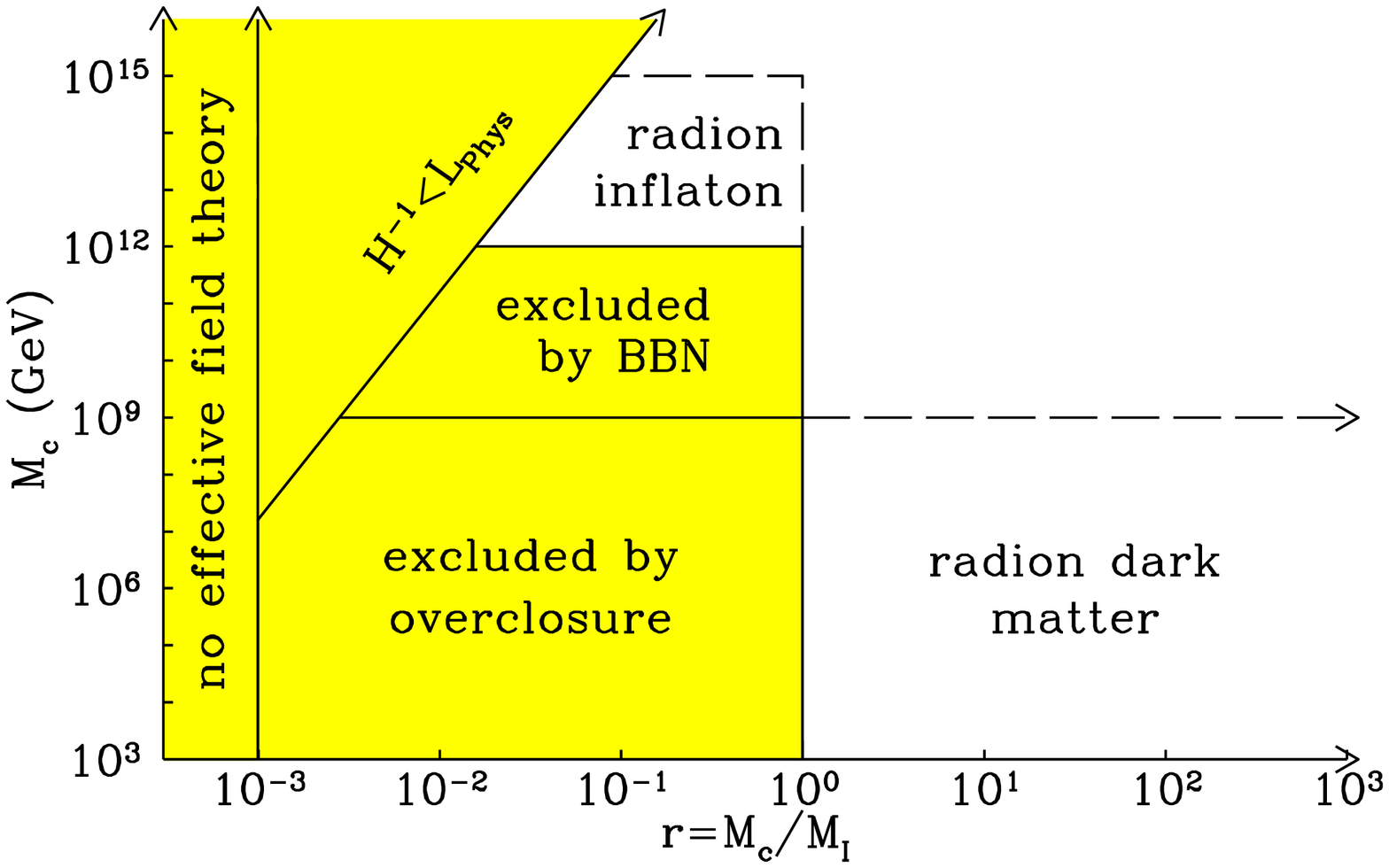,width=15cm}
{A summary of the cosmological significance in various regions of
radion model parameter space. $M_c$ is the compactification scale and $M_I$
is the scale of the potential of the scalar field responsible for inflation.
We note that the application of
effective field theory breaks down in the region $M_I\gtrsim 10^3M_C$, as 
well as the region where the
Hubble radius is smaller than the size of the extra dimension.
\label{fig:summary} }

We have written the scale of the radion potential as $M_c$ and used a
naturalness argument to identify $L \sim M_c^{-1}$, without specifying
the compactification dynamics.  It is interesting to compare this with
existing stabilization models, and the radion masses which result.  In
Ref.\ \cite{Ponton:2001hq}, it was shown that casimir effects will
stabilize the extra dimension (and in fact, are always present for a
compact space). This leads to a potential which is natural in our
sense, and an estimate of the radion mass similar to
Eq.~(\ref{radionmass}) \cite{Chacko:2002sb}.  An alternative (in 6
dimensions) is to invoke a bulk cosmological constant and a set of
gravitating $3$-branes, balanced against quantum effects from KK mode
gravitons \cite{Sundrum:1998ns}.  This also leads naturally to the
relation between the scale of the potential (characterized, {\em
i.e.,} by the bulk cosmological constant) and $L$ which we have
assumed.  Unlike the TeV$^{-1}$ (or smaller) extra dimensions we consider, the
aim of those works was to generate large (mm) size extra dimensions,
for which the radion is too light, in conflict with precision
measurements of gravity.  In order to generate acceptable large
extra-dimension phenomenology, one may introduce a bulk $U(1)$ gauge
theory and stabilize with its trapped magnetic flux
\cite{Sundrum:1998ns,Carroll:2001ih}.  An additional parameter allows
one to tune the radion potential such that the simple relation between
the radion mass and $M_c$ is disturbed, with the free parameter given
by the $U(1)$ coupling. Another way to avoid the natural relation 
$m_r\sim M_c^2/m_{Pl}$ was suggested in \cite{Arkani-Hamed:1999dz} 
where the stabilizing potential depends only logarithmically on the radion field so that a
large hierarchy between $M_c$ and $R^{-1}$ can be generated without fine-tuning. The idea
works for an even number of dimensions only since it relies on sets of two transverse
dimensions (for the logarithmic dependence). In addition, bulk supersymmetry is needed to
protect against the generation of a bulk cosmological constant which would induce power law
corrections to the effective potential for the radion and spoil the picture with logarithmic
potentials.

Concerning the radion coupling to an inflaton (or any scalar field),
the result is an interesting inflationary dynamics through which the
radion may be pushed to dominate the energy density of the universe
and play the role of the inflaton. Given the scale $M_I$ which
characterizes the potential of another scalar field, the cosmological
evolution of the universe depends crucially on the ratio $r=M_c/M_I$
where $M_c$ is the compactification scale. The coupling to the
inflaton will generally push the radion VEV away from its low energy
minimum, in way strongly reminiscent of the coherent production
mechanism of the axion. Models with low compactification scales (less
than about $10^9$ GeV) possess radions which are stable on
cosmological time scales, for which there are strong overclosure
constraints. When inflation ends, energy density is stored in radion
oscillations which can easily dominate the energy density of the
universe at late times.  This strong moduli problem requires that, for
models of conventional inflation, the inflation scale must be below
the compactification scale $r \gtrsim 1$.

An alternative is the possibility that the radion could account for the dark
matter required by cosmological measurements. Radion dark matter is probably
challenging to detect directly, though evidence might be available from
the gamma-ray sprectrum (through the eventual, rare radion decay) or
from isocurvature perturbations in the CMB.  

We have examined the possibility that the radion could play the role
of either the curvaton or modulon, and find that generally it is not
suitable.  It typically does not have large enough density
fluctuations to be either a curvaton (unless it is already the
inflaton) or a modulon.  Finally, we have explored the possibility
that the radion could decay after nucleosynthesis, and examined the
stringent bounds which result.  These bounds require both a much
smaller abundance from coherent production and require a low
reheat temperature to avoid thermal production mechanisms.

\subsection{TeV Extra Dimensions}

Models with TeV$^{-1}$ extra dimensions have attracted much attention
lately. Model builders in particle theory have made use of TeV$^{-1}$
extra dimensions to suggest new mechanisms for breaking symmetries,
naturally generating small numbers, and addressing the hierarchy
problem
\cite{Arkani-Hamed:1999dc,Arkani-Hamed:2000hv,Antoniadis:1998sd} using
orbifold boundary conditions. The natural scale for which these
mechanisms to operate is the natural scale of physics beyond the SM.
Thus, compactification should be at the TeV scale\footnote{We note 
that because the radion potential is protected by general
covariance from large quantum corrections, the choice of $M_c$ is stable
with respect to $M_*$, the higher dimensional Planck scale, and thus is
technically natural.  However, the
usual hierarchy problem, $M_{EW}$ compared to $M_*$, remains somewhat
mysterious in TeV extra dimensions (however, see
\cite{Arkani-Hamed:2000hv,Antoniadis:1998sd}).}.
Thus
 far, little
has been said\footnote{An observation related to the cosmology of TeV
extra dimensions was made in \cite{KorthalsAltes:2001et} where it was
found that there is a domain-wall problem in models with bulk $SU(N)$
gauge fields and fundamental brane matter if the reheat temperature is
larger than $M_c$.}  about the cosmology of these models, aside from
the possibility of Kaluza--Klein dark matter, which is only an option
in the special case of universal extra dimensions.
 
In the present work, we have derived additional and 
stronger constraints from the radion physics. We showed that
TeV$^{-1}$ extra dimensions generally have an overclosure problem,
because the radion is effectively stable and easily dominates
the energy density of the universe at late times.
The radion energy density can be sufficiently damped provided 
the inflation scale $M_I$ is at a TeV. 
The radion being stable, reheating would have to come from a 
different field which would imprint its fluctuations into the radiation. 

To summarize, in models with TeV$^{-1}$ extra dimensions:
\begin{itemize}
\item The radion is effectively stable.
\item To avoid overclosure of the universe by radions, either the inflationary
model must be rethought, or the inflation scale must be below the TeV
scale {\em i.e.,} $r\gtrsim 1$. In this case, $V_{\phi}$ is
subdominant during slow-roll and $\phi$ cannot be the inflaton.  There
is a narrow window of parameter space in which the radion can account
for the observed abundance of dark matter.
\item Inflation model building is challenging: 
A reheat temperature of order 100 GeV is preferred for baryogenesis
and also for dark matter (unless the radion is dark matter---the LSP
or LKP dark matter candidates actually require $T_{RH}\sim$
100 GeV). For $T_{RH}\sim$ TeV$\sim M_I$, this necessitates a very
efficient reheating, and almost certainly preheating.
\end{itemize}
In the last years, particle physics model building has moved into theories
with TeV extra dimensions. We have explored the cosmological
challenges associated with these models, and by seeing where constraints
arise, define the properties that a solution of these challenges should
have. One way out of these constraints points towards non-toroidal compactifications, like,
for instance, warped geometries \cite{Randall:1999ee,Goldberger:1999uk,Csaki:2000zn} 
or internal compact hyperbolic spaces for which $m_r\sim R^{-1}$ \cite{Kaloper:2000jb}.
  The future of extra-dimensional cosmology is both challenging and
exciting.

\acknowledgments
G.S. thanks C.~Armend\'ariz-Pic\'on for useful discussions and T.T. is
grateful for conversations with E.~Pont\'{o}n and to the SLAC theory group
for hospitality while part of this work was undertaken.  This work is
supported in part by the US Department of Energy, High Energy Physics
Division, under contract W-31-109-Eng-38, by the David and Lucile
Packard Foundation, and by NASA grant NAG5-10842.



\end{document}